\documentclass[12pt,onecolumn,draftcls]{IEEEtran}
\usepackage[dvips]{graphicx}
\usepackage{amscd}
\usepackage{amsmath,epsfig}
\usepackage{amssymb}
\usepackage{amsmath}
\usepackage{cases} 
\usepackage{amssymb,amsmath,cite,color}
\usepackage{epsfig}
\usepackage{slashbox}
\usepackage{bm}
\usepackage{amsthm}

\newcommand{\bb}{\mathbf{b}}

\newcommand{\be}{\mathbf{e}}

\newcommand{\bg}{\mathbf{g}}
\newcommand{\bp}{\mathbf{p}}

\newcommand{\bc}{\mathbf{c}}

\newcommand{\bI}{\mathbf{I}}
\newcommand{\bw}{\mathbf{w}}
\newcommand{\bW}{\mathbf{W}}
\newcommand{\bv}{\mathbf{v}}
\newcommand{\bx}{\mathbf{x}}

\newcommand{\bd}{\mathbf{d}}

\newcommand{\bA}{\mathbf{A}}
\newcommand{\bB}{\mathbf{A}}

\renewcommand{\frac}{\dfrac}

\newcommand{\SI}{{\mbox{SINR}}}
\newcommand{\T}{{{T}}}

\newcommand{\I}{{\cal I}}

\newcommand{\J}{{\cal J}}
\newcommand{\K}{{\cal K}}
\renewcommand{\S}{\cal S}

\newcommand{\st}{{\text{s.t.}}}

\newtheorem{dingyi}{Definition~}
\newtheorem{dingli}{Theorem~}
\newtheorem{yinli}{Lemma~}
\newtheorem{tuilun}{Corollary~}

\begin{document}

\title{{Joint Power and Admission Control: Non-Convex $L_q$ Approximation and An Effective Polynomial Time Deflation Approach}
\author{Ya-Feng Liu, Yu-Hong Dai, and Shiqian Ma}
\thanks{
Part of this work was presented in the IEEE International Conference on Acoustics, Speech,
and Signal Processing (ICASSP), Vancouver, Canada, May 26--31,
2013\cite{q-norm}.}
 \thanks{Y.-F.~Liu and Y.-H. Dai are with the State Key Laboratory
of Scientific and Engineering Computing, Institute of Computational
Mathematics and Scientific/Engineering Computing, Academy of
Mathematics and Systems Science, Chinese Academy of Sciences,
Beijing, 100190, China (e-mail: yafliu@lsec.cc.ac.cn;
dyh@lsec.cc.ac.cn).}
\thanks{S.~Ma is with the Department of Systems Engineering and Engineering Management, The Chinese University of Hong Kong,
Shatin, Hong Kong (e-mail:
 sqma@se.cuhk.edu.hk).}
 }

 \maketitle

 \begin{abstract}
     \boldmath In an interference limited network, joint power and admission control (JPAC) aims at supporting a maximum number of links at their specified signal to interference plus noise ratio (SINR) targets while using minimum total transmission power. Various convex approximation deflation approaches have been developed for the JPAC problem. In this paper, we propose an effective polynomial time \emph{non-convex} approximation deflation approach for solving the problem. {The approach} is based on the non-convex $\ell_q~(0<q<1)$ approximation of an equivalent sparse $\ell_0$ reformulation of the JPAC problem. We show that, for any instance of the JPAC problem, there exists {a} $\bar q\in(0,1)$ such that {it can be exactly solved by solving its $\ell_q$ approximation problem with any $q\in(0, \bar q]$.} 
     {We also show that finding the global solution of the $\ell_q$ approximation problem is NP-hard.} Then, we propose a potential reduction interior-point algorithm, which can return an $\epsilon$-KKT solution of the NP-hard $\ell_q$ approximation problem in polynomial time. The returned solution can be used to check the simultaneous supportability of all links in the network and to guide an iterative link removal procedure, resulting in the polynomial time non-convex approximation deflation approach for the JPAC problem. Numerical simulations show that the proposed approach outperforms the existing convex approximation approaches in terms of the number of supported links and the total transmission power, particularly exhibiting a quite good performance in selecting which subset of links to support.

%
%

     \end{abstract}
\section{Introduction}
Joint power and admission control (JPAC) has been recognized as an effective tool for interference
management in cellular, ad hoc, and cognitive underlay wireless
networks for more than two decades {\cite{convex_approximation,region,removals,Simple,ex2,ex3,admission_Bambos,ex1,bbound,SPAWC,msp,scheduling2,powerx,scheduling,varying,survey,decentralized,admission_luo,distributed,ACM,S1,S2,throughput,q-norm,beamformingscheduling,performance,downlink_power,fully1,fully2,zander2,robust}}.
The goal of JPAC is to support a maximum number of links at their specified signal to interference plus noise ratio (SINR) targets while using minimum total transmission power when all links in the interference limited network {cannot} be simultaneously supported. JPAC {can not only} determine which interfering links must be turned off and rescheduled along orthogonal resource
dimensions (such as time, space, or frequency slots), {but also alleviate the difficulties
of the convergence} of stand-alone power control algorithms. For example, a longstanding issue associated with {the Foschini-Miljanic} algorithm \cite{Simple} is that{,} it does not converge when the preselected SINR levels are {infeasible}. In this case, a JPAC approach must be adopted to determine which links {to} be removed.

\subsection{Related Work}
The JPAC problem can be solved to global optimality by checking the
simultaneous supportability of every subset of links. However, the
computational complexity of this enumeration approach grows
exponentially with the total number of links. {Another globally optimal algorithm, which is based on the branch and bound strategy, is given in
\cite{bbound}.} Theoretically, the problem is {shown} to be NP-hard to solve (to global optimality) {and} to approximate ({to constant factor of global optimality}) \cite{convex_approximation,removals,msp}. In recent years, {various} convex approximation based heuristics
algorithms \cite{convex_approximation,region,removals,Simple,ex2,ex3,admission_Bambos,ex1,bbound,msp,SPAWC,scheduling2,powerx,scheduling,varying,survey,decentralized,admission_luo,distributed,S1,S2,ACM,throughput,q-norm,beamformingscheduling,performance,robust} have been proposed for {the} problem, {since convex optimization problems, such as linear program (LP), second-order cone program (SOCP), and semidefinite {program (SDP)}, are relatively easy to solve\footnote{For any convex
optimization problem and any $\epsilon>0$, the ellipsoid
algorithm can find an $\epsilon$-optimal
solution (i.e., a feasible solution whose objective value is within
$\epsilon$ from being globally optimal) with a complexity that is
polynomial in the problem dimension and $\log(1/\epsilon)$ \cite{com3}.}.}

{Assuming} perfect channel state information (CSI), Ref.~\cite{convex_approximation} proposed the so-called linear programming deflation (LPD) algorithm. Instead of solving the original NP-hard problem directly, the LPD algorithm
solves an appropriate LP approximation of the original problem at each iteration and {uses} its solution to guide the removal of interfering links. The removal procedure is {repeated} until all the remaining links in
the network are simultaneously supportable. {In \cite{msp}, the JPAC problem is shown to be equivalent to a sparse $\ell_0$ minimization problem and then its $\ell_1$ convex relaxation is used to derive {an LP}, which is different from the one in \cite{convex_approximation}. Again, the solution to the derived LP is used to guide an iterative link removal procedure (deflation), leading to an efficient new linear programming deflation (NLPD) algorithm.} {Another convex approximation based heuristics
algorithm is proposed in \cite{region}. Assuming the same SINR target for each link, the link that
results in the largest increase in the achievable SINR is removed at each iteration
until all the remaining links in the network are simultaneously supportable.} To determine the removed link, {a large number of extreme eigenvalue problems\footnote{Extreme eigenvalue problems are SDP representable.} need to be solved at each iteration, making the removal procedure computationally expensive.} To reduce the computational complexity, the above idea is approximately implemented in the Algorithm II-B
\cite{region}.
~Similar convex approximation deflation ideas {were used} in \cite{admission_luo,decentralized} to solve the joint beamforming and admission control problem for the cellular downlink network, where at each iteration {an SDP} needs to be solved to determine the link to be removed.

Under the imperfect CSI assumption, JPAC {has been} studied in \cite{convex_approximation,robust,SPAWC}. In \cite{convex_approximation}, the authors considered the worst-case robust JPAC problem with bounded channel estimation errors. The key there is that the relaxed LP with bounded uncertainty can be equivalently rewritten as {an SOCP}. The overall approximation algorithm remains similar to LPD for the case of the perfect CSI, except that the SOCP formulation is used to carry out power control and its solution is used to check whether links are simultaneously supportable in the worst case. Ref.~\cite{robust} studied the JPAC problem under the assumption of the channel distribution information (CDI), and formulated the JPAC problem as a chance (probabilistic) constrained program, where each link's SINR outage probability is enforced to be less than or equal to a specified tolerance. To circumvent the difficulty of the chance SINR constraint, Ref.~\cite{robust} employed the sample (scenario) approximation scheme to convert the chance constraints into finitely many simple linear constraints. Then, the sample approximation of the chance SINR constrained JPAC problem is reformulated as a group sparse minimization problem and approximated by {an} SOCP. The solution of the SOCP approximation problem can be used to check the simultaneous supportability of all links in the network and to guide an iterative link removal procedure.

\subsection{Our Contribution}
This paper considers the JPAC problem under the perfect CSI assumption. We remark that similar techniques can be used {for} the case where the CSI is not perfectly known. As mentioned above, most existing algorithms on JPAC are based on (successive) convex approximations. The main contribution of this paper is {to} propose an effective polynomial time non-convex approximation deflation approach for solving the JPAC problem. {To our knowledge, this is the first {approach} that solves the JPAC problem by (successive) non-convex approximations.} The key idea is to approximate the sparse $\ell_0$ minimization reformulation of the JPAC problem by the non-convex $\ell_q$ minimization problem with $q\in(0,1)$ instead of the convex $\ell_1$ minimization problem as in\cite{convex_approximation,msp}, and {to design a polynomial time algorithm for computing an $\epsilon$-KKT solution (its definition will be given later) of the non-convex $\ell_q$ minimization problem for any given $\epsilon>0$}.
The main results of this paper are {summarized} as follows.

%

\begin{itemize}
  \item We show that the non-convex $\ell_q$ minimization approximation problem shares the same solution with the $\ell_0$ minimization problem if $q\in (0, \bar{q}],$ where $\bar{q}$ is some value in $(0, 1)$. We also give an example of the JPAC problem, showing that the solution to its non-convex $\ell_q$ minimization approximation problem with any $q\in (0,1)$ solves the original problem while its convex $\ell_1$ minimization approximation problem fails to do so. We therefore show that the $\ell_q$ minimization problem with $q\in(0,1)$ approximates the $\ell_0$ minimization JPAC problem better than the $\ell_1$ minimization problem.

  \item We show that, for any $q\in (0,1),$ the $\ell_q$ minimization approximation problem is NP-hard. The proof is based on a polynomial time transformation from the partition problem. The complexity result suggests that there is no polynomial time algorithm which can solve the $\ell_q$ minimization approximation problem to global optimality (unless P$=$NP).
  \item We reformulate the $\ell_q$ minimization approximation problem and develop a potential reduction interior-point algorithm for solving its equivalent reformulation. We show that, for any given $\epsilon>0,$ {the potential reduction algorithm can return an $\epsilon$-KKT solution of the reformulated problem in polynomial time.} {The obtained $\epsilon$-KKT solution can be used to check the simultaneous supportability of all links in the network and to guide an iterative link removal procedure, resulting in the polynomial time non-convex approximation deflation approach for the JPAC problem. Simulation results show that the proposed approach significantly outperforms the existing convex approximation algorithms\cite{msp,region,convex_approximation}.}

\end{itemize}

%
%
%
\subsection{Notations}
We adopt the following notations in this paper. We denote the index
set $\{1,2,\ldots,K\}$ by ${\K}$. Lowercase boldface and uppercase
boldface are used for vectors and matrices, respectively. For a
given vector $\bx,$ the notations $\max\{\bx\},$ $[\bx]_k$ and $\|\bx\|_q^q:=\sum_{k}|[\bx]_k|^q~(0\leq q\leq1)$ stand
for its maximum entry, its $k$-th entry, and its $\ell_q$ norm\footnote{Strictly speaking, $\|\bx\|_q^q$ with $0\leq q<1$ is not a norm, since it does not satisfy the triangle inequality. However, we still call it $\ell_q$ norm for convenience in this paper.}, respectively. In particular, when $q=0,$ $\|\bx\|_0$ stands for the number of nonzero entries in $\bx.$
~For any subset $\I\subseteq {\cal K}$, we use $\bA_\I$ to denote
the matrix formed by the rows of $\bA$ indexed by $\I$. For a vector
$\bx$, the notation $\bx_\I$ is similarly defined. Moreover, for any
$\J\subseteq {\cal K}$, the notation $\bA_{\I,\J}$ will denote the
submatrix of $\bA$ obtained by taking the rows and columns of $\bA$
indexed by $\I$ and $\J$ respectively. 
The spectral radius of a matrix $\bB$ is denoted by $\rho(\bB).$
Finally, we use $\be$ to represent the vector with all components
being one and $\bI$ to represent the identity matrix of an
appropriate size,
respectively.

\section{{System Model, Sparse Formulation, and NLPD Algorithm}}
Consider a $K$-link (a link corresponds to a transmitter-receiver
pair) interference channel with channel gains $g_{k,j}\geq0$ (from
transmitter $j$ to receiver $k$), noise power $\eta_k>0,$ SINR
target $\gamma_k>0,$ and power budget $\bar p_k>0$ for $k, j\in
{\K}:=\{1,2,\ldots,K\}$. Denote the power allocation vector by
$\bp=(p_1,p_2,\ldots,p_K)^\T$ and the power budget vector by $\bar
\bp=(\bar p_1,\bar p_2,\ldots,\bar p_K)^\T$. Treating interference
as noise, we can write the SINR at the $k$-th receiver as
\begin{equation}\nonumber\label{scale}\displaystyle
\SI_k=\frac{g_{k,k}p_k}{\eta_k+\displaystyle\sum_{j\neq
k}g_{k,j}p_j},\quad\forall~k\in\K.\end{equation}

{To some extent,} the JPAC problem can be
formulated as a two-stage optimization problem. The
first stage
maximizes the number of admitted links:
\begin{equation}\label{MSP1}
\begin{array}{ll}
\displaystyle \max_{{\bp,\,{\cal S}}} & \displaystyle |\cal S| \\
[5pt] \mbox{s.t.} & \displaystyle
\SI_k\geq \gamma_k,~k\in\cal S\subseteq\cal K,\\
&\bm{0} \leq \bp\leq \bar\bp.
\end{array}
\end{equation}
{The optimal solution ${{\cal S}_0}$ of problem \eqref{MSP1}, which may not be unique, is called \emph{maximum admissible set}.}
The second stage minimizes the total transmission power required to support the admitted links in ${{\cal S}_0}$:
\begin{equation}\label{MSP2}
\begin{array}{cl}
 \displaystyle\min_{\left\{p_k\right\}_{k\in{\cal S}_0}} & \sum_{k\in {{\cal S}_0}} p_k \\
  \mbox{s.t.} &  \SI_k\geq \gamma_k,~k\in{{\cal S}_0},\\
&\displaystyle 0\leq p_k\leq \bar p_k,~
k\in{\cal S}_0.
\end{array}
\end{equation}
Due to the special choice of ${\cal S}_0,$ power control problem
\eqref{MSP2} is feasible and can be efficiently solved by the
Foschini-Miljanic algorithm \cite{Simple}.

{The two-stage JPAC problem \eqref{MSP1} and \eqref{MSP2} is reformulated as a
single-stage sparse $\ell_0$~minimization problem in \cite{msp}, which is based on a normalized channel. Next, we first introduce the channel normalization and then the sparse formulation of the JPAC problem. Denote the normalized power allocation vector by
$\bx=\left(x_1,x_2,\ldots,x_K\right)^T$ with
$x_k={p_k}/{\bar p_k}$, and the normalized noise vector by
${\bb}=\left(b_1,b_2,\ldots,b_K\right )^T$ with
  $b_k={\left(\gamma_k\eta_k\right)}/{\left(g_{k,k}\bar p_k\right)}>0$.
The normalized channel matrix is denoted by $\bA\in {\mathbb R}^{K\times
K}$ with the $(k,j)$-th entry
\begin{equation*}\label{A}
a_{k,j}=\left\{\begin{array}{cl}
1,&\text{if~}k=j;\\
\displaystyle - \frac{\gamma_kg_{k,j}\bar p_j}{g_{k,k}\bar p_k},&\text{if~}k\neq j.
\end{array}
\right.
\end{equation*}
The JPAC problem can be reformulated as a single-stage sparse $\ell_0$~minimization problem as follows
\begin{equation}\label{sparse2}
\begin{array}{cl}
\displaystyle \min_{\bx} & \|\bb-\bA\bx\|_0+\alpha\,\bar\bp^T\bx \\
\mbox{s.t.} & \displaystyle \mathbf{0}\leq \bx\leq \be,
\end{array}
\end{equation}
where $\alpha$ is a parameter satisfying
\begin{equation}\label{alpha1}0< \alpha<\alpha_1:=1/{\be^T\bar\bp}.\end{equation}
Notice that the formulation \eqref{sparse2} is capable of finding the maximum admissible set with minimum total
transmission power and hence is superior to the two-stage formulation \eqref{MSP1} and \eqref{MSP2} in case of multiple maximum admissible sets.}

{The basic idea of the NLPD algorithm in \cite{msp} is to
update the power and check whether all links can be supported. {If not, drop one link from the network and update the power again.} This process is repeated
until all the remaining links are supported. More specifically, the NLPD algorithm checks whether all links in the network can be simultaneously supported by solving the $\ell_1$ convex approximation of the $\ell_0$ minimization problem \eqref{sparse2} 
\begin{equation}\label{ll1}
\begin{array}{cl}
\displaystyle \min_{\bx} & \|\bb-\bA\bx\|_1+\alpha\bar\bp^T\bx \\
\mbox{s.t.} & \displaystyle \mathbf{0}\leq \bx\leq \be,
\end{array}
\end{equation}which is equivalent to the following LP (see Theorem 2 in \cite{msp})
\begin{equation}\label{linear}
  \begin{array}{cl}
   \displaystyle \min_{\bx} & \be^\T\left(\bb-\bA\bx\right)+\alpha\,\bar \bp^T\bx\\
\mbox{s.t.} &\bb-\bA\bx\geq\mathbf{0},\\
                  &\mathbf{0}\leq \bx\leq \be.
  \end{array}
\end{equation}If all links in the network cannot be simultaneously supported, the NLPD algorithm drops the link 
\begin{equation}\label{remove2}k_0=\arg\max_{k\in\cal K}\left\{\sum_{j\neq k}\left(|a_{k,j}|\left[\bb-\bA\bx\right]_j+|a_{j,k}|\left[\bb-\bA\bx\right]_k\right)\right\}.\end{equation}
To accelerate the deflation process, an easy-to-check necessary condition 
\begin{equation}\label{necessary}
\left(\bm{\mu}^+\right)^T\be-\left(\bm{\mu}^-+\be\right)^T\bb\geq0
\end{equation}
 for all links in the
network to be simultaneously supported is also derived in
\cite{msp}, where $\bm{\mu}^+=\max\left\{\bm{\mu},\mathbf{0}\right\},$
$\bm{\mu}^-=\max\left\{-\bm{\mu},\mathbf{0}\right\},$ and
$\bm{\mu}=\bA^\T\be$. The necessary condition allows to iteratively
remove strong interfering links from the network. In particular, the link \begin{equation}\label{sum}
k_0=\arg\max_{k\in\cal K}\left\{\sum_{j\neq
k}|a_{k,j}|+\sum_{j\neq k}|a_{j,k}|+b_k\right\}
\end{equation} is iteratively removed in the NLPD algorithm until \eqref{necessary} becomes true.
}

The complete description of the NLPD algorithm is given as
follows.\vspace{0cm}\begin{center}
\framebox{
\begin{minipage}{12.3cm}
\flushright
\begin{minipage}{12.3cm}
\centerline{\bf Algorithm 1: The NLPD Algorithm}\vspace{0.05cm} \textbf{Step 1.}
Initialization: Input data
$\left(\bA,\bb,\bar\bp\right).$\\[2.5pt]
\textbf{Step 2.} Preprocessing: Remove link $k_0$ iteratively
according to \eqref{sum} until condition \eqref{necessary} holds
true.\\[2.5pt]
\textbf{Step 3.} Power control: Solve problem \eqref{ll1}; check
whether all links are supported: if yes, go to \textbf{Step
5}; else go to \textbf{Step 4}. \\[2.5pt]
\textbf{Step 4.} Admission control: Remove link  $k_0$ according to
\eqref{remove2}, set ${\K}={\K} \setminus \left\{k_0\right\},$ and go to
\textbf{Step 3}.\\[2.5pt]
\textbf{Step 5.} Postprocessing: Check the removed links for possible admission.
\end{minipage}
\end{minipage}
}
\end{center}

\section{{A Non-Convex $L_q$ Approximation Deflation Approach for JPAC}}
{In this section, we develop a polynomial time non-convex $\ell_q~(0<q<1)$ approximation deflation algorithm for the JPAC problem. The motivation for developing such a non-convex approximation deflation algorithm is that the $\ell_q$ minimization problem with $q\in(0,1)$ should perform better than the $\ell_1$ minimization problem in approximating the $\ell_0$ minimization problem \eqref{sparse2} and thus the deflation algorithm based on non-convex $\ell_q$ approximations should have a better performance than the NLPD algorithm, which is based on convex $\ell_1$ approximations.

In the following, we first analyze exact recovery of the $\ell_q$ minimization approximation in solving the $\ell_0$ minimization problem in Section \ref{q-recovery}. Then, we prove in Section \ref{q-nphard} that the $\ell_q$ minimization problem with any $q\in(0,1)$ is NP-hard. In Section \ref{q-ipalgorithm}, we develop a polynomial time interior-point algorithm for approximately solving the $\ell_q$ minimization problem. The approximate solution of the $\ell_q$ minimization problem can be used to check the simultaneous supportability of all links in the network and to guide an iterative link removal procedure, thus resulting in the polynomial time non-convex approximation deflation approach for the JPAC problem in Section \ref{q-deflation}.}

%

\subsection{{Exact Recovery of Non-Convex $L_q$ Approximation}}\label{q-recovery}
The sparse $\ell_0$ minimization problem \eqref{sparse2} is \emph{successively} approximated by the $\ell_1$~minimization problem \eqref{ll1} 
in the NLPD algorithm. Intuitively, the $\ell_q$ minimization problem with $0<q<1$,
\begin{equation}\label{lp}
\begin{array}{cl}
\displaystyle \min_{\bx} & \|\bb-\bA\bx\|_q^q+\alpha\,\bar\bp^T\bx \\
\mbox{s.t.} & \displaystyle \mathbf{0}\leq
\bx\leq \be
\end{array}
\end{equation}
should approximate \eqref{sparse2} ``better'' than \eqref{ll1}. To provide such {an evidence}, we give the following lemma.  

\begin{yinli}\label{balancing}
  For any $q\in[0,1],$ problem \eqref{lp} is equivalent to
    \begin{equation}\label{qnorm-r}
\begin{array}{cl}
\displaystyle \min_{\bx} & \displaystyle \|\bb-\bA\bx\|_q^q + \alpha \bar\bp^T\bx \\
[3pt] \st & \bA\bx\leq \bb,\\
& \displaystyle \mathbf{0}\leq \bx\leq \be.
\end{array}
\end{equation}
\end{yinli}
The above lemma can be verified in a similar way as the proof of Theorem 2 in \cite{msp} {and a detailed proof is provided in Section I of \cite{technicalreport}}.
Based on Lemma \ref{balancing}, we can show the following result (see Appendix \ref{app-recover}
for its proof).
\begin{dingli}\label{thm-recover}
  For any given instance of problem \eqref{sparse2}, there exists $\bar q>0$ (depending on $\bA,\bb,\alpha,\bar\bp$) such that when $q\in (0,\bar q],$ {any} global solution to problem \eqref{lp} is one of the global solutions to problem \eqref{sparse2}. 
\end{dingli}

 Theorem \ref{thm-recover} states that the $\ell_q$ minimization problem \eqref{lp} shares the same solution with the $\ell_0$ minimization problem \eqref{sparse2} if the parameter $q$ (depending on $\bA,\bb,\alpha,\bar\bp$) is chosen to be sufficiently small. In general, the $\ell_1$~minimization problem \eqref{ll1} does not enjoy this exact recovery property, which is in sharp contrast to the results in \cite{tao} and \cite{donoho}.

It is shown in \cite{tao} that the problem of minimizing $\|\bA\bx-\bb\|_1$
is equivalent to the problem of minimizing $\|\bA\bx-\bb\|_0$ with high probability if the vector $\bA\bx-\bb$ at the true solution $\bx^*$ is sparse, where $\bA\in\mathbb{R}^{m\times n}$ and $m>n$, and if the entries of the matrix $\bA$ are independent and identically distributed (i.i.d.) Gaussian.
The reason why the $\ell_1$~minimization problem \eqref{lp} fails to recover the solution of problem \eqref{sparse2} is that the two assumptions required in
  \cite{tao} do not hold true for problem \eqref{sparse2}. Specifically, the vector $\bA\bx-\bb$ may not be sparse even at the optimal power allocation vector $\bx^*$. This depends on whether the (normalized) channel is strongly interfered or not. {More importantly}, the matrix $\bA$ in \eqref{sparse2} is a square matrix and has a special structure; i.e., all diagonal entries are one and all non-diagonal entries are non-positive.

Next we give an example to illustrate the advantage of the use of $\ell_q$ norm with $0<q<1$ over the use of {$\ell_1$ norm} to approximate problem \eqref{sparse2}. Suppose $\bA,\bb,\bar\bp$ in \eqref{sparse2} are given as follows:
$$\bA=\left(
                                        \begin{array}{ccc}
                                          1 & 0 & -1 \\
                                          0 & 1 & -1 \\
                                          -1 & -1 & 1 \\
                                        \end{array}
                                      \right),~\bb=0.5 \be,~\bar\bp=\be.
$$ {It can be shown that the optimal solution to the sparse optimization problem \eqref{sparse2} is $$\bx^*=(0.5,0.5,0)^T$$ if the parameter $\alpha$ is chosen satisfying $0<\alpha<1/3$ (cf. \eqref{alpha1}).} 
We can also obtain the solutions to problems \eqref{ll1} and \eqref{lp}.
\begin{itemize}
  \item By writing the KKT optimality conditions, we can check that $\bx=0$ is the unique global minimizer of problem \eqref{ll1} with any $\alpha\geq 0$.
  \item Lemma \ref{balancing} implies that problem \eqref{lp} is equivalent to 
\begin{equation}\label{0.5}
\begin{array}{cl}
\displaystyle \min_{x_1,x_2,x_3} & \displaystyle {(0.5-x_1+x_3)^q}+{(0.5-x_2+x_3)^q}+{(0.5+x_1+x_2-x_3)^q} + \alpha(x_1+x_2+x_3) \\
[5pt] \mbox{s.t.} & 0.5-x_1+x_3\geq 0,\\
&0.5-x_2+x_3\geq 0,\\
&0.5+x_1+x_2-x_3\geq 0,\\
&0\leq x_1,x_2,x_3\leq 1.
\end{array}
\end{equation} For any given $q\in(0,1),$ define \begin{equation}\label{ccc}\bar\alpha_q:={\min\left\{1+(0.5)^q, 2^q\right\}-(1.5)^q}>0.\end{equation}
It can be checked (although tedious) that, as long as $\alpha$ in problem \eqref{0.5} is chosen such that $0<\alpha\leq \bar \alpha_q,$~
the unique global minimizer of problem \eqref{0.5} is $\bx^*$; {see Section II of \cite{technicalreport}}.
\end{itemize}

We remark that, for a given instance of problem \eqref{sparse2}, it is generally not easy to determine $\bar q$ in Theorem \ref{thm-recover}. {However, our simulation results in Section \ref{subsec-simulation} show that it is generally not very small for small networks.} In practice, we could set the parameter $q$ in problem \eqref{lp} to be a constant in $(0,1)$ ({more on the choice of the parameter $q$ will be discussed in Section \ref{sec-simulation}).}
~Therefore, the solution to problem \eqref{lp} might not be able to solve the $\ell_0$~minimization problem \eqref{sparse2}. This is the reason why we do not just use the $\ell_q$~minimization \eqref{lp} to approximate problem \eqref{sparse2}, but instead employ a deflation technique to successively approximate problem \eqref{sparse2} in our proposed algorithm below.

\subsection{Complexity Analysis of $L_q$~Minimization \eqref{lp}}\label{q-nphard}
{Roughly} speaking, convex optimization problems are relatively easy to solve, while non-convex
optimization problems are difficult to solve. However, not all non-convex problems
are computationally intractable since the lack of convexity may be due to an inappropriate
formulation. In fact, many non-convex optimization
problems admit a convex reformulation; see \cite{complexity,coordinated,simo,Coordition,precoding,RQ,multi-user} for some examples. Therefore, convexity
is useful but unreliable to evaluate the computational intractability
of an optimization problem. A more robust tool is the computational
complexity theory \cite{Complexitybook,combinatorial}.
%

In this subsection, we show that problem \eqref{lp} is NP-hard for any given $q\in(0,1)$. The NP-hardness proof is based on a polynomial time
transformation from the partition problem: given a set of $N$ positive integers $s_1,s_2,\ldots,s_N,$ determine whether there exists a subset $\S$ of $\left\{1,2,\ldots,N\right\}$ such that $$\sum_{n\in \S}s_n=\sum_{n\notin \S}s_n=\frac{1}{2}\sum_{n=1}^Ns_n.$$ The partition problem is known to be NP-complete\cite{Complexitybook}.

\begin{dingli}\label{hardness}
  For any given $0< q<1,$ the $\ell_q$~minimization problem \eqref{lp} is NP-hard.
\end{dingli}

The proof of Theorem \ref{hardness} is relegated to Appendix \ref{app-hardness}. Theorem \ref{hardness} suggests that there is no efficient algorithm which can solve problem \eqref{lp} to global optimality in polynomial time (unless P$=$NP), and finding an approximate solution for it is more realistic in practice.

{We remark that some related problems \begin{equation}\label{ye}
 \begin{array}{rl}
\displaystyle \min_{\bx} & \displaystyle \left\|\bx\right\|_q^q \\
                     \mbox{s.t.} & \displaystyle \bA\bx=\bb\end{array}
\end{equation} and
\begin{equation}\label{l2lq}
  \min_{\bx} \displaystyle \left\|\bA\bx-\bb\right\|_p^p+\lambda\left\|\bx\right\|_q^q
\end{equation}
are shown to be NP-hard in \cite{ye} and \cite{chen}, where $p\in[1,\infty)$ and $q\in(0,1)$. However, these NP-hardness results cannot imply our result in Theorem \ref{hardness}, since the considered problems are different. A key difference between our problem (10) and problems \eqref{ye} and \eqref{l2lq} is that problem (10) (equivalent to \eqref{qnorm-r}) tries to find an $\bx$ such that the number of positive entries of the vector $\bb-\bA\bx$ is as small as possible while problems \eqref{ye} and \eqref{l2lq} try to find a solution such that the number of nonzero entries of $\bx$ is as small as possible. Moreover, the matrix $\bA$ and the vector $\bb$ in (10) have special structures, i.e., all diagonal entries of $\bA$ are one, all non-diagonal entries of $\bA$ are non-positive, and all entries of $\bb$ are positive. This restriction on $\bA$ and $\bb$ makes it more technical and intricate to show the NP-hardness of problem \eqref{lp} compared to show that of problems \eqref{ye} and \eqref{l2lq} for general $\bA$ and $\bb$ in \cite{ye} and \cite{chen}.
}

\subsection{A Polynomial Time Potential Reduction Algorithm for Problem \eqref{lp}}\label{q-ipalgorithm}
In this subsection, we develop a polynomial time potential reduction interior-point algorithm for solving problem \eqref{lp}.
Based on Lemma \ref{balancing}, by introducing slack variables, we see that problem \eqref{lp} can be equivalently formulated as
\begin{equation}\label{pnorm1}
\begin{array}{cl}
\displaystyle \min_{\bw} & \displaystyle f(\bw):=\tilde\bc^T\bw_1+\|\bw_2\|_q^q \\
[5pt] \mbox{s.t.} & \tilde \bA \bw=\tilde \bb,\\
& \displaystyle \bw\geq \mathbf{0},
\end{array}
\end{equation}where $$\tilde \bA=\left(
    \begin{array}{ccc}
      \bA & \bI & \mathbf{0} \\
      \bI & \mathbf{0} & \bI \\
    \end{array}
  \right)\in\mathbb{R}^{2K\times 3K}, \tilde \bb=\left(
                         \begin{array}{c}
                           \bb \\
                           \be \\
                         \end{array}
                       \right)\in\mathbb{R}^{2K},~\tilde\bc=\alpha\bar\bp\in\mathbb{R}^{K},~\bw=\left(
                                    \begin{array}{c}
                                      \bw_1 \\
                                      \bw_2 \\
                                      \bw_3 \\
                                    \end{array}
                                  \right)\in\mathbb{R}^{3K}.$$
{We extend the potential reduction algorithm in \cite{ye,ye2} to solve problem \eqref{pnorm1}} to obtain one of its $\epsilon$-KKT points (the definition of the $\epsilon$-KKT point shall be given later). 
It can be shown that the potential reduction interior-point algorithm returns an $\epsilon$-KKT point 
of problem \eqref{pnorm1} in no more than $O\left(\left(\frac{K}{\min\left\{\epsilon,q\right\}}\right)\log\left(\frac{1}{\epsilon}\right)\right)$ iterations.

Before going into {the} details, we first give a high level preview of the proposed algorithm. The two basic ingredients of the potential reduction interior-point algorithm is the potential function (cf. \eqref{potential}) and the update rule (cf. \eqref{preduction}). The potential function measures the progress of the algorithm, and the update rule guides to compute the next iterate based on the current one. More specifically, the next iterate is chosen as the feasible point that achieves the maximum potential reduction. The algorithm is terminated either when the potential function is below some threshold (cf. \eqref{pupper}) or when the potential reduction (cf. \eqref{perreduction}) is smaller than a constant. In the former case, the algorithm returns an $\epsilon$-optimal solution of problem \eqref{pnorm1}, and in the latter case an $\epsilon$-KKT point of problem \eqref{pnorm1}. Moreover, the polynomial time convergence of the algorithm can be guaranteed by showing that the value of the potential function is decreased by at least a constant at each iteration and the potential function is {bounded above and below by some thresholds}.
%
%

\textbf{Definition of $\epsilon$-KKT point.} {Since $f(\bw)$ is not differentiable when some entries of $\bw_2$ are zero, the common definitions of the KKT point do not apply to problem \eqref{pnorm1}. We thus define a weaker concept, the so-called $\epsilon$-KKT point (or $\epsilon$-KKT solution) for problem \eqref{pnorm1}, in a similar way as done in \cite{ye,ye2,KKT}.}
~Suppose that $\bw^*$ is a local minimizer of problem \eqref{pnorm1}, and define ${\S}=\left\{j~|~[\bw^*_2]_j>0\right\}.$ 
Then $\bw_{\S}^*=(\bw_1^*; [\bw_2^*]_{\S}; \bw_3^*)$ should be a local minimizer of problem
\begin{equation}\label{reducepnorm}
\begin{array}{cl}
\displaystyle \min_{\bw_{\S}} & \displaystyle \tilde\bc^T\bw_1+\|[\bw_2]_{\S}\|_q^q\\
[5pt] \mbox{s.t.} & \tilde \bA_{\S} \bw_{\S}=\tilde \bb,\\
& \displaystyle \bw_{\S}\geq \mathbf{0},
\end{array}
\end{equation}
where $\tilde \bA_{\S}=\left(
    \begin{array}{ccc}
      \bA & \bI_{\S}^T & \mathbf{0} \\
      \bI & \mathbf{0} & \bI \\
    \end{array}
  \right)\in\mathbb{R}^{(2K)\times (2K+|{\S}|)}.$ 
The KKT condition of problem \eqref{reducepnorm} is: there exists a Lagrange multiplier vector $\bm{\lambda}^*\in\mathbb{R}^{2K}$ such that \begin{equation}\label{dual}\nabla_{\bw_{\S}} f(\bw^*)-\tilde \bA_{\S}^T\bm{\lambda}^*\geq \mathbf{0}\end{equation} and 
$$(\nabla_{\bw_{\S}} f(\bw^*)-\tilde \bA_{\S}^T\bm{\lambda}^*)^T\bw^*_{\S}=0.$$ 
Notice that given $\bm{\lambda}^*,$ if $\left[\bw^*\right]_n=0$ for $n-K\notin\S,$ we have
 $$q\left[\bw^*\right]_n^{q} -\left[\tilde \bA^T\bm{\lambda}^*\right]_n\left[\bw^*\right]_n=0.$$
 Therefore, the $\epsilon$-KKT point of problem \eqref{pnorm1} can be defined as follows.

  \begin{dingyi}\label{eKKT}$\bw^*$ is called an $\epsilon$-KKT point of problem \eqref{pnorm1} if
  \begin{itemize}
    \item [(a)] it is feasible;
    \item [(b)] there exists $\bm{\lambda}^*$ such that \eqref{dual} holds true; and
    \item [(c)] the complementarity gap
    \begin{equation}\label{gap}\displaystyle\frac{\displaystyle\sum_{n=1}^{3K}\left(q\left[\bw^*\right]_n^{q} -\left[\tilde \bA^T\bm{\lambda}^*\right]_n\left[\bw^*\right]_n\right)}{\bar f-\underline f}\leq \epsilon,
    \end{equation}
 where $\bar f$ and $\underline f$ are upper and lower bounds on the objective value of problem \eqref{pnorm1}, respectively.
  \end{itemize}
  In addition, $\bw^*$ is called an $\epsilon$-optimal solution to problem \eqref{pnorm1} if $f(\bw^*)\leq \epsilon.$
 \end{dingyi}

 It is worthwhile remarking that if $\epsilon=0$ in \eqref{gap}, the above definition reduces to the definition of the KKT point {of problem \eqref{reducepnorm}.}
 For simplicity, we set $\underline f=0$ in \eqref{gap} in this paper, since the objective function of problem \eqref{pnorm1} is always nonnegative.


\textbf{Potential Function.} 
For any given {strictly feasible} $\bw$, define the following potential function
\begin{equation}\label{potential}
  \phi(\bw)= \rho\log (f(\bw))-\sum_{k=1}^{3K}\log([\bw]_k),
\end{equation}where $\rho$ is a parameter to be specified later.

\begin{yinli}\label{lemma-lowerbound}
  Let $\epsilon>0$ and $\rho>\frac{K}{q}$ be fixed. Suppose that $\bw$ is strictly feasible and satisfies \begin{equation}\label{pupper}\phi(\bw)\leq \left(\rho-\frac{K}{q}\right)\log(\epsilon)+\frac{K}{q}\log \left(K\right)+K\log(4).\end{equation} Then $\bw$ is an $\epsilon$-optimal solution to problem \eqref{pnorm1}.
\end{yinli}

{Lemma \ref{lemma-lowerbound} actually gives a lower bound of the potential function $\phi(\bw)$. Its proof can be found in Appendix \ref{app-potential}. In the following, we provide an upper bound for $\phi(\bw).$
Let $$\bw^0=\left(
                                    \begin{array}{c}
                                      \frac{\min\left\{\bb,\be\right\}}{2} \\[8pt]
                                      \bb-\frac{\bA\min\left\{\bb,\be\right\}}{2} \\[8pt]
                                      \be-\frac{\min\left\{\bb,\be\right\}}{2} \\
                                    \end{array}
                                  \right).$$ It can be verified that $\bw^0$ in the above is an interior point of problem (16) by the use of the special structures of $\bA$ and $\bb.$ Since the potential function values are decreasing at each iteration $\bw^t$ of the potential reduction algorithm (see ``\textbf{Update Rule}'' further ahead), it follows that
                                  \begin{equation}\label{upperbound-w}
                                    \phi(\bw^t)\leq \phi(\bw^0),~\forall~t\geq0,
                                  \end{equation}where $\phi(\bw^0)$ is a constant depending only on the problem inputs $\bA,\bb,\bar\bp$ and $\alpha.$ 
}

\textbf{Update Rule.} 
Consider one iteration update from $\bw$ to $\bw^+$ by minimizing the potential reduction $\phi(\bw^+)-\phi(\bw).$ Suppose that $\bw^+=\bw+\bd>\mathbf{0},$ where $\bd$ satisfies
$\tilde \bA \bd=\mathbf{0}.$ From the concavity of $\log(f(\bw)),$ we have
\begin{equation}\label{concave}\log(f(\bw^+))-\log(f(\bw))\leq \frac{1}{f(\bw)}\nabla f(\bw)^T \bd.\end{equation} On the other hand, we have the following standard lemma\cite[Theorem 9.5]{Bertsimas}.

\begin{yinli}\label{standlemma}
  Let $\bW=\text{Diag}(\bw).$ Suppose that $\|\bW^{-1}\bd\|\leq \beta<1.$ Then, we have
  $$-\sum_{k=1}^{3K}\log([\bw^+]_k)+\sum_{k=1}^{3K}\log([\bw]_k)\leq -\be^T\bW^{-1}\bd+\frac{\beta^2}{2(1-\beta)}.$$
\end{yinli}\noindent It is worthwhile remarking that if $\|\bW^{-1}\bd\|\leq \beta<1,$ then $\bw^+=\bw+\bd>\mathbf{0}.$
By combining \eqref{concave} and Lemma \ref{standlemma}, we have
\begin{align}
  \phi(\bw^+)-\phi(\bw)\leq \left(\frac{\rho}{f(\bw)}\nabla f(\bw)^T\bW-\be^T\right)\bW^{-1}\bd+\frac{\beta^2}{2(1-\beta)}.\label{reduction}
\end{align} Let $\tilde \bd=\bW^{-1}\bd.$ To achieve the maximum potential reduction, one can solve the following problem
\begin{equation}\label{preduction}
\begin{array}{cl}
\displaystyle \min_{\tilde \bd} & \displaystyle \bv^T\tilde \bd \\
[5pt] \mbox{s.t.} & \tilde \bA \bW\tilde \bd=\mathbf{0},\\
& \displaystyle \|\tilde \bd\|^2\leq\beta^2,
\end{array}
\end{equation}where $$\bv=\frac{\rho}{f(\bw)}\bW\nabla f(\bw)-\be.$$ Problem \eqref{preduction} is simply a projection problem.
The minimal value of problem \eqref{preduction} is
\begin{equation}\label{perreduction}-\beta\|g(\bw)\|,\end{equation} and the solution to problem \eqref{preduction} is $\tilde \bd=\frac{\beta}{\|g(\bw)\|}g(\bw),$ where
$$g(\bw)=\be-\frac{\rho}{f(\bw)}\bW\left(\nabla f(\bw)-\tilde \bA^T\bm{\lambda}\right),$$and
$$\bm{\lambda}=\left(\tilde \bA\bW^2\tilde \bA^T\right)^{-1}\tilde \bA\bW\left(\bW\nabla f(\bw)-\frac{f(\bw)}{\rho}\be\right).$$

The polynomial time complexity of the above potential reduction interior-point algorithm is summarized in the following theorem.
Its proof is relegated to Appendix \ref{app-polynomialtimecomplexity}.
\begin{dingli}\label{theoremK}
  The potential reduction interior-point algorithm returns an $\epsilon$-KKT point or $\epsilon$-optimal solution
  of problem \eqref{pnorm1} (equivalent to problem \eqref{lp}) in no more than $O\left(\left(\frac{K}{\min\left\{\epsilon,q\right\}}\right)\log\left(\frac{1}{\epsilon}\right)\right)$ iterations.
\end{dingli}
{Since one projection problem {in the form of} \eqref{preduction} needs to be solved at each iteration of the potential reduction algorithm, and the complexity of solving problem \eqref{preduction} is $O(K^{3})$, we immediately have the following corollary.
\begin{tuilun}\label{corollaryK}
  The potential reduction interior-point algorithm returns an $\epsilon$-KKT point or $\epsilon$-optimal solution
  of problem \eqref{pnorm1} (equivalent to problem \eqref{lp}) in no more than $O\left(\left(\frac{K^4}{\min\left\{\epsilon,q\right\}}\right)\log\left(\frac{1}{\epsilon}\right)\right)$ operations.
\end{tuilun}}

One may ask why we restrict ourselves to use interior-point algorithms for solving problem \eqref{pnorm1} (equivalent to problem \eqref{lp}). The reasons are the following.
First, the objective function of problem \eqref{pnorm1} is \emph{differentiable} in the interior feasible region. Moreover, we are actually interested in finding a feasible $\bw$ such that $\bw_2$ is as sparse as possible; if we start from a $\bw$, some entries of $\bw_2$ are already zero, then it is very hard to make it nonzero. 
In contrast, if we start from an interior point, the interior-point algorithm may generate a sequence of interior points that bypasses solutions with the wrong zero supporting set and converges to the true one. This is exactly the idea of the interior-point algorithm developed in \cite{ye2} for the non-convex quadratic programming.

{To further improve the solution quality, we propose to run the above potential reduction algorithm multiple times to solve problem \eqref{pnorm1} and pick the best one among (potentially different) returned $\epsilon$-KKT solutions, where at each time the potential reduction algorithm is initialized with a randomly generated interior point. Thanks to the special structure of $\bA$ and $\bb,$ it can be verified that the random point} {\begin{equation}\label{initialization}\bw\left(\bm{\xi}\right)=\left(
                                    \begin{array}{c}
                                      {\bm{\xi}\circ\min\left\{\bb,\be\right\}} \\[8pt]
                                      \bb-\bA\left({\bm{\xi}\circ\min\left\{\bb,\be\right\}}\right) \\[8pt]
                                      \be-{\bm{\xi}\circ\min\left\{\bb,\be\right\}} \\
                                    \end{array}
                                  \right)\end{equation}} {is an interior point to problem \eqref{pnorm1} with probability one, where each entry of $\bm{\xi}$ obeys the uniform distribution in the interval $[0,1]$ and $\circ$ is the Hadamard product operator. 
                                  }

\subsection{A Polynomial Time Non-Convex Approximation Deflation Approach}\label{q-deflation}

The proposed $\ell_q$~minimization deflation ({LQMD}) algorithm, based on successive $\ell_q$ minimization approximations, is given as follows. {The key difference between the LQMD algorithm and the NLPD algorithm lies in the power control step (i.e., \textbf{Step 3}), albeit the framework of the two algorithms are the same.}

\begin{center}
\framebox{
\begin{minipage}{12.3cm}
\flushright
\begin{minipage}{12.3cm}
\centerline{\bf Algorithm 2: The LQMD Algorithm}\vspace{0.05cm} \textbf{Step 1.}
Initialization: Input data
$\left(\bA,\bb,\bar \bp\right),$ $q\in(0,1),$ and {positive integer $N.$}\\[2.5pt]
\textbf{Step 2.} Preprocessing: Remove link $k_0$ iteratively
according to \eqref{sum} until condition \eqref{necessary} holds
true.\\[2.5pt]
\textbf{Step 3.} Power control: {Compute the parameter $\alpha$ by \eqref{optalpha2} and run the potential reduction algorithm with $N$ randomly generated initial points \eqref{initialization} to solve problem \eqref{lp}}; check
whether all links are supported: if yes, go to \textbf{Step 5};
else
go to \textbf{Step 4}. \\[2.5pt]
\textbf{Step 4.} Admission control: Remove link  $k_0$ according to
\eqref{remove2}, set ${\K}={\K}\setminus\left\{k_0\right\},$ and go to
\textbf{Step 3}.\\[2.5pt]
\textbf{Step 5.} Postprocessing: Check the removed links for possible admission.
\end{minipage}
\end{minipage}
}
\end{center}

{Two remarks on the LQMD algorithm are in order. First, the parameter $\alpha$ in $\ell_q$ minimization \eqref{lp} is computed by
\begin{equation}\label{optalpha2}\alpha=\left\{\!\!\!\!
  \begin{array}{rll}
  &c_1\alpha_1, & \text{if}~\rho(\bI-\bA)\geq 1;\\
  &\min\left\{c_2\alpha_1,\,c_3\alpha_2\right\}, & \text{if}~\rho(\bI-\bA)< 1,
  \end{array}\right.
\end{equation}where $c_3>c_2,$ $0<c_1, c_2<1$ are three constants. In the above, $\alpha_1$ is determined by the equivalence between problem \eqref{sparse2} and the joint problem \eqref{MSP1} and \eqref{MSP2} (cf. \eqref{alpha1}), and $\alpha_2$ is determined by the so-called ``Never-Over-Removal'' property (cf. (22) of \cite{msp}). 
Second, the LQMD has a polynomial time worst-case complexity, which is \begin{equation}\label{complexitydeflation}O\left(\left(\frac{NK^5}{\min\left\{\epsilon,q\right\}}\right)\log\left(\frac{1}{\epsilon}\right)\right).\end{equation} This is because that at most $K$ links will be dropped and the complexity of dropping one link needs solving problem \eqref{lp} $N$ times in the LQMD algorithm. Combing this with Corollary \ref{corollaryK}, we immediately obtain the complexity result in \eqref{complexitydeflation}.}

\section{Numerical Simulations}\label{sec-simulation}
{In this section, we carry out two sets of numerical experiments to illustrate the effectiveness of the non-convex $\ell_q$ approximation \eqref{lp} and the LQMD algorithm (Algorithm 2), respectively. We employ the number of supported links and the total transmission power as the comparison metrics to compare different approximations and algorithms. In our simulations, the parameters in \eqref{optalpha2} are set to be $c_1=c_2=0.2$~and~$c_3=4$. 
}

We generate the same channel parameters as in \cite{convex_approximation} in our simulations; i.e., each
transmitter's location obeys the uniform distribution over a $2$
Km~$\times$~$2$ Km square and the location of its corresponding
receiver is uniformly generated in a disc with radius $400$ m;
channel gains are given by \begin{equation}\label{gain}g_{k,j}=\frac{1}{d_{k,j}^4},~\forall~k,\,j\in\K,\end{equation}
where $d_{k,j}$ is the Euclidean distance from the link of
transmitter $j$ to the link of receiver $k.$ Each link's SINR target
is set to be $\gamma_k=2~\text{dB}~(\forall~k\in\K)$ and the noise
power is set to be $\eta_k=-90~\text{dBm}~(\forall~k\in\K)$. The
power budget of the link of transmitter $k$ is
\begin{equation}\label{budget}\bar p_k=2p_k^{\min},~\forall~k\in\K,\end{equation}  where $p_k^{\min}$ is
the minimum power needed for link $k$ to meet its SINR requirement
in the absence of any interference from other links.

{\begin{center}
\framebox{
\begin{minipage}{12.3cm}
\flushright
\begin{minipage}{12.3cm}
\centerline{\bf Algorithm 3: A Heuristic Algorithm for Computing $\bar q$}\vspace{0.05cm} \textbf{Step 1.}
Input data $\left(\bA,\bb,\bar \bp\right),$ positive integer $N,$ and nonempty set $\mathcal{Q}.$  \\[2.5pt]
\textbf{Step 2.} Compute the parameter $\alpha$ by \eqref{optalpha2}. Use the enumeration method to solve problem \eqref{sparse2} and denote the global solution by $\bx^{\ast}$.\\[2.5pt]
\textbf{Step 3.} If $\mathcal{Q}$ is empty, set $\bar q=0,$ return \textsf{failure} and terminate the algorithm; else pick the largest $q\in\mathcal{Q}$ and run the potential reduction algorithm with $N$ randomly generated initial points \eqref{initialization} to solve problem \eqref{lp}. Denote the best point by $\bx_q^N.$  \\[2.5pt]
\textbf{Step 4.} If $\bx_q^N=\bx^{\ast},$ set $\bar q=q,$ return \textsf{success} and terminate the algorithm; else set $\mathcal{Q}=\mathcal{Q}\setminus \left\{q\right\}$ and go to
\textbf{Step 3}.
\end{minipage}
\end{minipage}
}
\end{center}}

\subsection{Non-Convex $L_q$ versus Convex $L_1$ Approximations}\label{subsec-simulation}

{In this subsection, we do numerical simulations in two small networks where there are $K=5$ and $K=10$ links, respectively.

We first test how small $q$ needs to be for the $\ell_q$ minimization problem \eqref{lp} to exactly recover the global solution of problem \eqref{sparse2}. We propose Algorithm 3 to heuristically compute $\bar q$ in Theorem \ref{thm-recover}. In Algorithm 3, the parameters $N$ and $\mathcal{Q}$ are set to be 
\begin{equation}\label{NQ}N=100,~\mathcal{Q}=\left\{0.01,0.02,\ldots,1\right\}.\end{equation}



Fig. \ref{barqq} depicts the computed $\bar q$ of $100$ random channel realizations where $K=5$ and $K=10.$
It shows that the parameter $\bar q$ required in Theorem \ref{thm-recover} is heavily problem-dependent. Fortunately, it is generally not very small, i.e., the average $\bar q$ of the $100$ random channel realizations for the two networks where $K=5$ and $K=10$ are $0.5364$ and $0.1951,$ respectively. Fig. \ref{barqq} also suggests that a smaller $\bar q$ tends to be required as the number of total links in the network becomes large.

 Among the above $100$ random channel realizations, the $\ell_q$ minimization problem \eqref{lp}, with the parameter $q$ being judiciously chosen by Algorithm 3, successfully finds the global solution of problem \eqref{sparse2} $99$ times and $82$ times  for the two networks where $K=5$ and $K=10$, respectively. This is consistent with our analysis in Theorem \ref{thm-recover}. The simulation results also show good performance of the potential reduction algorithm with multiple random initializations in finding the global solution of problem \eqref{lp}. As we can see, there are some instances that problem \eqref{lp} fails to find the global solution of problem \eqref{sparse2}. The possible reasons are: a) the required $\bar q$ in Theorem \ref{thm-recover} for these instances might be less than $0.01,$ which is the smallest value we test in our simulations (cf. $\mathcal{Q}$ in \eqref{NQ}); and/or b) the potential reduction algorithm (even with $100$ random initializations) does not find the global solution of problem \eqref{lp}, which is NP-hard as shown in Theorem \ref{hardness}.





\begin{figure}[!t]
     \centering
     \centerline{\includegraphics[width=10.2cm]{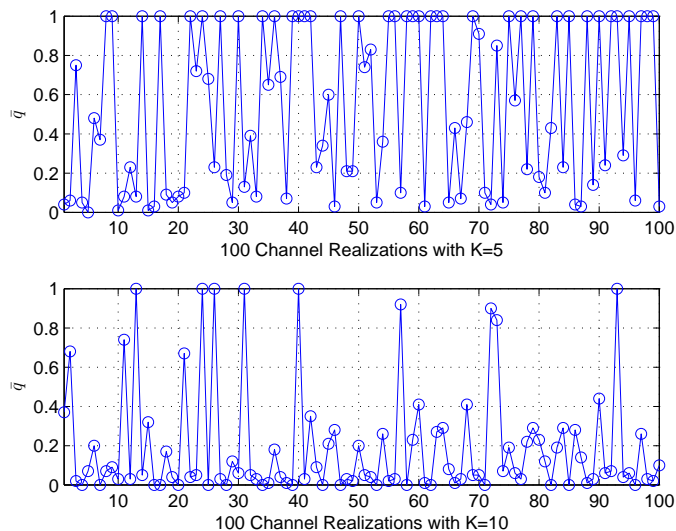}}
     \caption{{The computed $\bar q$ of $100$ channel realizations with $K=5$ and $K=10.$}}
     \label{barqq}
     \end{figure}


We now compare the performance of the $\ell_q$ minimization problem \eqref{lp} with $q$ being fixed to be $0.1$ and the $\ell_1$ minimization problem \eqref{ll1} in approximating the $\ell_0$ minimization problem \eqref{sparse2}. The corresponding $\ell_q$ minimization problem \eqref{lp} is solved again by running the potential reduction algorithm with $100$ random initializations and the corresponding $\ell_1$ minimization problem \eqref{ll1} is solved by using the {simplex} method to solve its equivalent LP reformulation \eqref{linear}. The global solution obtained by ``brute-force'' enumeration is used as benchmark. Simulation results are summarized in Table \ref{tableoneshot} and Fig. \ref{oneshotk10}. Table \ref{tableoneshot} reports the performance comparison of $\ell_{0.1}$ and $\ell_1$ approximations in terms of average number of supported links, average total transmission power, and percentage of finding the global solution of problem \eqref{sparse2} in the $100$ random channel realizations. Fig. \ref{oneshotk10} illustrates the number of supported links and the total transmission power of $10$ random channel realizations for the network where $K=10$.

\begin{table*}[!ht]
\caption{\textsc{{Statistics of $L_{0.1}$ and $L_1$ approximations for $100$ Random Channel Realizations}}} \label{tableoneshot} \centering
\begin{tabular}{cccccccc}
\hline \hline\multicolumn{1}{c}{} & \multicolumn{3}{c}{$K=5$}  & \multicolumn{3}{c}{$K=10$}\\
\cline{2-7}
  &Links&Power&Percentage&Links&Power&Percentage\\\hline
 Benchmark&    4.05 &43.71 & 100\%  &  6.46 & 60.57& 100\%   \\
 $\ell_{0.1}$&    4.05 &51.36 & 69\%  &  6.32 & 67.86& 26\%   \\
 $\ell_{1}$&    3.86 &49.90 & 47\%  &  5.28 & 51.94& 3\%   \\\hline
\hline
\end{tabular}
\end{table*}

\begin{figure}[!t]
     \centering
     \centerline{\includegraphics[width=10.2cm]{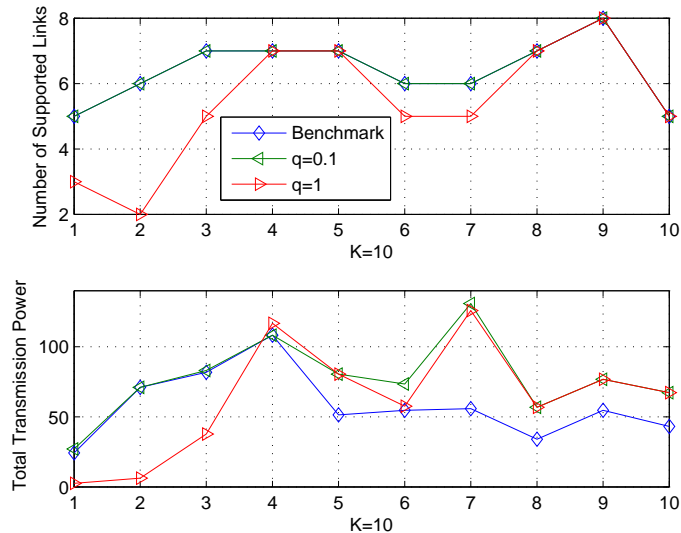}}
     \caption{{Comparison of $\ell_{0.1}$ and $\ell_1$ for $10$ random channel realizations for the network where $K=10$.}}
     \label{oneshotk10}
     \end{figure}

Table \ref{tableoneshot} shows that the $\ell_{0.1}$ minimization approximation significantly outperforms the $\ell_1$ minimization approximation in terms of the number of supported links.
~It can be observed from Table \ref{tableoneshot} that the $\ell_{0.1}$ minimization problem can find the maximum admissible set for all of the $100$ random channel realizations and successfully finds the maximum admissible set with minimum total transmission power with a percentage of $69\%$ for the case $K=5.$ For the case $K=10,$ the percentage of the $\ell_{0.1}$ minimization problem for finding the `optimal' maximum admissible set decreases to 26\%, but still it performs much better than the $\ell_1$ minimization problem in the sense that it supports $1.04$ more links than $\ell_1$ minimization in average; see Table \ref{tableoneshot}. Fig. \ref{oneshotk10} shows that the $\ell_{0.1}$ minimization problem finds the maximum admissible set for all these $10$ channel realizations, and successfully finds the `optimal' one for the first $4$ channel realizations. Therefore, the $\ell_q$ minimization problem \eqref{lp} indeed exhibits a significantly better capability in approximating the $\ell_0$ minimization problem \eqref{sparse2} compared to the $\ell_1$ minimization problem \eqref{ll1}.
}



\subsection{{Effectiveness of LQMD}}
{We present some numerical simulation results to evaluate the effectiveness of the proposed LQMD algorithm in this part. We set $N=5$ in the LQMD algorithm. 
All figures in this subsection are obtained by averaging over $200$ Monte-Carlo runs.

We first test whether the performance of the LQMD algorithm is sensitive to the choice of the parameter $q.$ Figs. \ref{sensitivityuser} and \ref{sensitivitypower} plot the performance comparison of the LQMD algorithm with different choices of the parameter $q.$ 
More specifically, for each fixed $K$ and $q\in\left\{0.1,0.3,0.5,0.7,0.9\right\},$ we use the LQMD algorithm to solve $200$ randomly generated JPAC problems, and denote the average number of supported links by $N(K,q).$ Each point in Fig. \ref{sensitivityuser} denotes 
$$D(K,q):=N(K,q)-\max_{q\in\left\{0.1,0.3,0.5,0.7,0.9\right\}} N(K,q).$$

\begin{figure}[!t]
     \centering
     \centerline{\includegraphics[width=10.2cm]{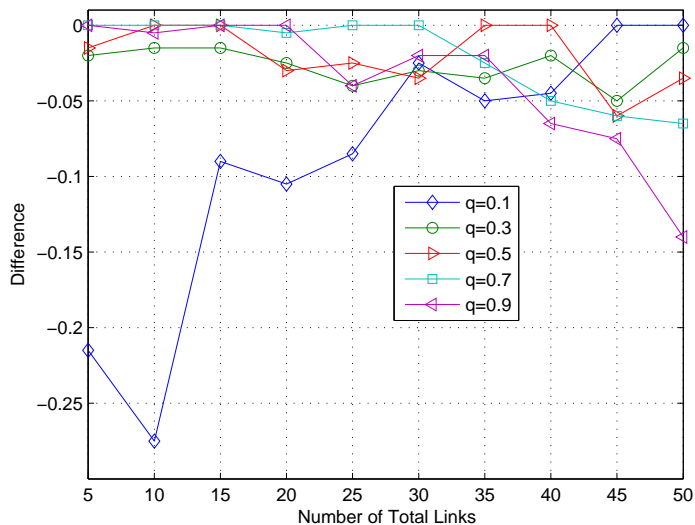}}
     \caption{{The difference of average number of supported links by the LQMD algorithm with different $q$
and the one by the LQMD algorithm with best $q\in\left\{0.1,0.3,0.5,0.7,0.9\right\}$ versus the number of total links.}}
     \label{sensitivityuser}
     \end{figure}

It can be observed from Figs. \ref{sensitivityuser} and \ref{sensitivitypower} that the performance of the LQMD algorithm is somehow sensitive to the choice of the parameter $q,$ which is mainly due to our implementation of the LQMD algorithm. Theoretically, the parameter $q$ should be chosen as small as possible according to Theorem 1 (if the $\ell_q$ approximation problem can be solved to global optimality). However, as shown in Theorem 2, finding the global solution of the $\ell_q$ minimization problem is NP-hard for any $q\in(0,1).$ In our implementation of the LQMD algorithm, we run the potential reduction algorithm with $5$ random initializations to solve the $\ell_q$ minimization problem at affordable complexity. As can be seen from Fig. \ref{sensitivityuser}, the performance of the LQMD algorithm in terms of the number of supported links with $q=0.1$ is not as good as expected. The main reason for this is because the $\ell_q$ minimization problem in the LQMD algorithm is solved by the potential reduction algorithm where the number of random initializations is set to be $5$ and in this case the potential reduction algorithm will get stuck at a local minimizer of the $\ell_q$ minimization problem with a higher probability for a small $q$ compared to a large $q$. One may set the number of random initializations to be very large, then the $\ell_{q}$ minimization problem with $q=0.1$ will be solved to global optimality with a high probability and thus the corresponding LQMD algorithm will enjoy a good performance. However, this will lead to excessively high computational costs and make the corresponding LQMD algorithm impractical. 
On the other hand, a large $q$ is also not suitable for the LQMD algorithm, since $\ell_q$ minimization with a large $q$ cannot approximate the $\ell_0$ minimization problem as good as the one with a small $q.$ This can be clearly seen from Figs. \ref{sensitivityuser} and \ref{sensitivitypower}, where the performance of the LQMD algorithm with $q=0.7~\text{and}~q=0.9$ in terms of the number of supported links gradually deteriorate as the number of total links increases and the corresponding total transmission power is larger than that of $q=0.1,0.3,0.5$ in the whole range.

     \begin{figure}[!t]
     \centering
     \centerline{\includegraphics[width=10.2cm]{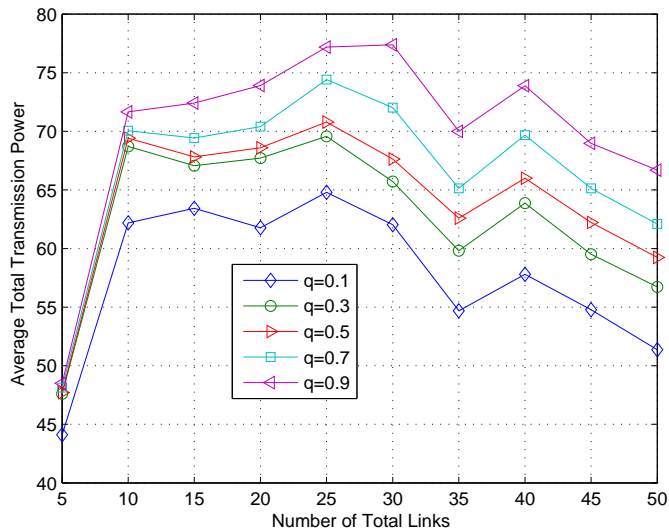}}
     \caption{{Average total transmission power by the LQMD algorithm with different $q$ versus the number of total links.}}
     \label{sensitivitypower}
     \end{figure}

The above simulation results and discussions provide us useful insights into the choice of the parameter $q,$ i.e., both small and large $q$ are not suitable for the LQMD algorithm due to either the practical implementation issue or the theoretical approximation issue and a median $q$ is preferred in the LQMD algorithm in terms of leveraging the implementation issue and enjoying a relatively good approximation property. Figs. \ref{sensitivityuser} and \ref{sensitivitypower} suggest that $q=0.3~\text{and}~q=0.5$ are the best ones in terms of the number of supported links and the total transmission power. We thus set $q=0.5$ in all of the following simulations.

We now compare the performance of the
proposed LQMD algorithm with that of the LPD
algorithm in \cite{convex_approximation}, the NLPD algorithm in \cite{msp}, and the Algorithm II-B in
\cite{region}, since all of them have been reported to have close-to-optimal performance in terms of the number of supported links. Figs.\ \ref{user1} to \ref{power1} plot the performance comparison
of aforementioned various admission and power control algorithms. Fig.\ \ref{user1} shows that the proposed LQMD algorithm and the NLPD algorithm can support more links than the other two algorithms (the LPD algorithm and the Algorithm II-B) over the whole range of the tested number of total links. Figs. \ref{user1} and \ref{power1} show that, compared to the LPD algorithm and the Algorithm II-B, the proposed LQMD algorithm can support more links with much less total transmission power.

\begin{figure}[!t]
     \centering
     \centerline{\includegraphics[width=10.2cm]{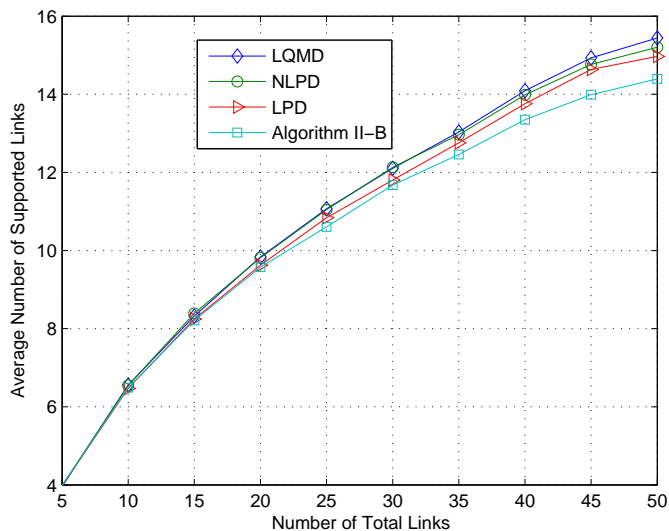}}
     \caption{{Average number of supported links versus the number of total
     links.}}
     \label{user1}
     \end{figure}
     \begin{figure}[!t]
     \centering
     \centerline{\includegraphics[width=10.2cm]{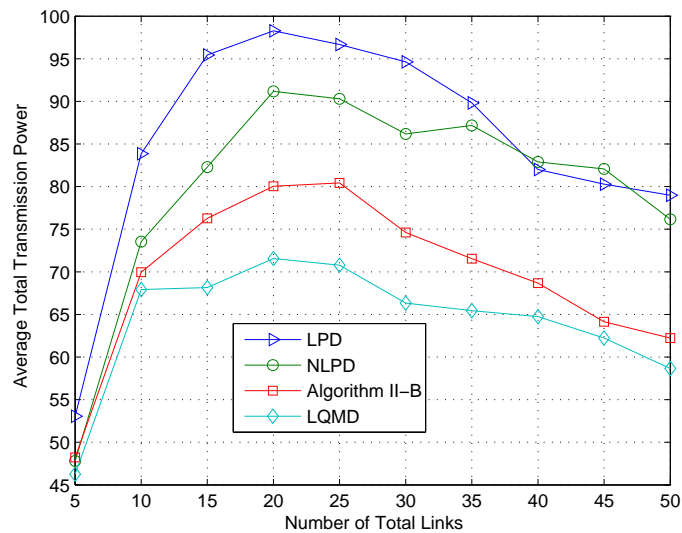}}
     \caption{{Average total transmission power versus the number of total
     links.}}
     \label{power1}
     \end{figure}

\begin{figure}[!t]
     \centering
     \centerline{\includegraphics[width=10.2cm]{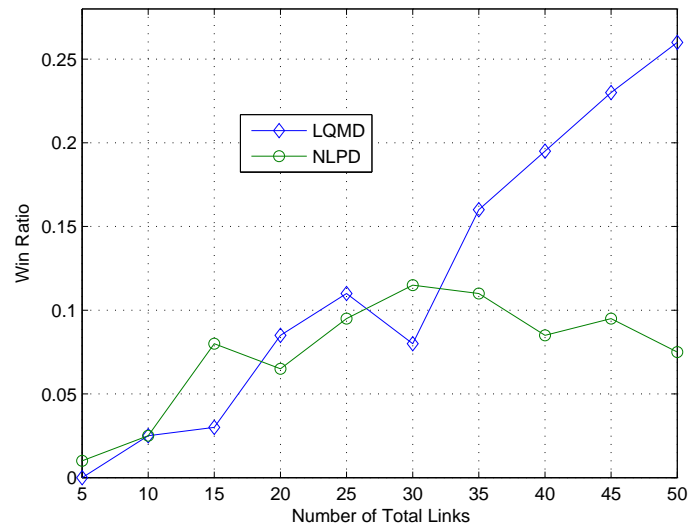}}
     \caption{{Win ratio comparison of LQMD and NLPD versus the number of total
     links.}}
     \label{Ratio}
     \end{figure}

\begin{figure}[!t]
     \centering
     \centerline{\includegraphics[width=10.2cm]{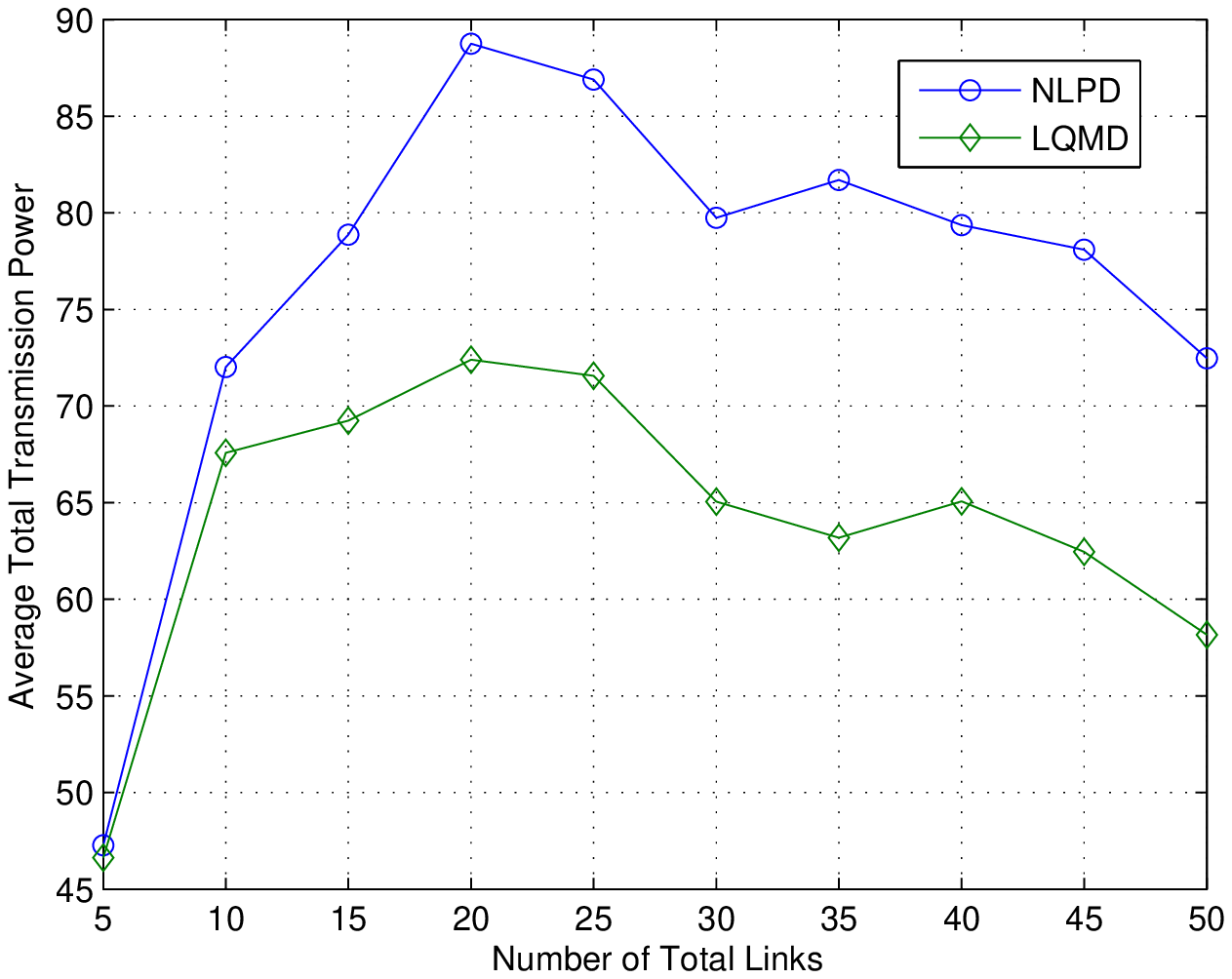}}
     \caption{{Average total transmission power comparison of NLPD and LQMD when the two algorithms find the admissible set with same cardinality versus the number of total links.}}
     \label{comNLPD}
     \end{figure}

Next, we focus on the performance comparison of the LQMD algorithm and the NLPD algorithm, since these two algorithms outperform the other two in terms of the number of supported links. The comparison results are presented in Figs. \ref{Ratio} and \ref{comNLPD}. The vertical axis ``Win Ratio'' in Fig. \ref{Ratio} shows the ratio of the number that the LQMD algorithm (the NLPD algorithm) winning the NLPD algorithm (the LQMD algorithm) to the total run number $200.$ Given an instance of the JPAC problem, the LQMD algorithm is said to win the NLPD algorithm if the former can support strictly more links than the latter for this instance. In a similar fashion, we can define that the NLPD algorithm wins the LQMD algorithm. It can be observed from Fig. \ref{Ratio} that the win ratio of the two algorithms are almost the same when the number of total links is less than or equal to $30,$\footnote{{In fact, it is impossible for the LQMD algorithm to achieve a large margin of the number of supported links over the NLPD algorithm for small networks, since it has been shown in \cite{msp} that the NLPD algorithm can achieve more than 98\% of global optimality (by ``brute force'' enumeration) in terms of the number of supported links when $K\leq18.$}} but the proposed LQMD algorithm wins the NLPD algorithm with a higher ratio when the number of total links is greater than $30.$ As depicted in Fig. \ref{Ratio}, when there are $K=50$ links in the network, the LQMD algorithm wins the NLPD algorithm $52$ times, while the NLPD algorithm wins the LQMD algorithm only $15$ times (among the total $200$ runs). The two algorithms find the admissible set with same cardinality for the remaining $133$ times. However, this does not mean that the two algorithms find the same admissible set in these cases. Fig. \ref{comNLPD} plots the average total transmission power when the two algorithms can support the same number of links, which demonstrates that the LQMD algorithm {is able to} select a ``better'' subset of links to support, and can use much less total transmission power to support the same number of links (compared to the NLPD algorithm). As the number of total links in the network increases, the LQMD algorithm saves more power. In a nutshell, the LQMD algorithm exhibits a substantially better performance (than the NLPD algorithm) in selecting which subset of links to support, and thus yields a much better total transmission power performance.
}




\begin{table*}[!ht]
\caption{{Ratios of Average Number of Supported Links and Total Transmission Power in Setup1 to That in Setup2.}}\label{table-ratio}\centering
 \begin{tabular}{|c|c|c|c|c|c|c|c|c|c|c|}
  \hline\hline
   \backslashbox{Ratio}{$K$} & 5 & 10 & 15 & 20 & 25 & 30 & 35 & 40 & 45 & 50 \\\hline
  Number of Supported Links & 0.9875 & 1.0085 & 0.9946 & 0.9919 & 1.0189 & 1.0017 & 0.9992 & 0.9975 & 1.0000 & 0.9834\\\hline
  Total Transmission Power & 3.6163 & 4.0054 & 4.0821 & 4.1320 & 3.9742 & 3.9698 & 3.9674 &4.2082 &4.2756 & 3.9979\\
  \hline\hline
\end{tabular}\end{table*}

{Finally, we test the performance of the LQMD algorithm in setups with different levels of interference. For convenience, we call the former simulation setup as Setup1. We decrease the distances between all transmitters and receivers by a factor of $0.707$ in Setup1 and call the obtained setup as Setup2. Notice that all the direct-link and cross-link channel gains and thus interference levels in Setup2 is $(1/0.707)^4\approx4$ times larger than that of Setup1 according to \eqref{gain}. 

Table \ref{table-ratio} summarizes the ratio of the average number of supported links and total transmission power in Setup1 to that in Setup2. It can be observed from it that the average number of supported links in both setups are roughly equal to each other, but the average total transmission power in Setup1 is approximately $4$ times as large as that in Setup2. This is because when the simulation setup is switched from Setup1 to Setup2, all the channel gains increase by a factor of $4$ and power budgets of all links decrease by a factor of $4$ (cf. \eqref{budget}).  As the channel gains are increased and power budgets are decreased by a same factor while the noise powers remain to be fixed, the number of supported links in problem \eqref{sparse2} remains unchanged. However, it brings a benefit of a 75\% reduction in the total transmission power, which is consistent with our engineering practice. Since the channel parameters in Setup1 and Setup2 are independently randomly generated, the ratios of average number of supported links and total transmission power in Setup1 to that in Setup2 in Table \ref{table-ratio} are approximately (but not exactly) $1$ and $4.$
}

\section{Conclusions}
In this paper, we have proposed a polynomial time non-convex approximation deflation approach for the NP-hard joint power and admission control (JPAC) problem. 
Different from the existing convex approximation approaches, the proposed one solves the JPAC problem by successive non-convex $\ell_q~(0<q<1)$~minimization approximations. We have shown exact recovery of the $\ell_q$~minimization problem, i.e., any global solution to the $\ell_q$~minimization problem is one of the global solutions to the JPAC problem as long as the parameter $q$ is chosen to be sufficiently small. We have also developed a polynomial time potential reduction interior-point algorithm for solving the $\ell_q$~minimization problem, which makes the proposed deflation approach enjoy a polynomial time worst-case complexity. Numerical simulations demonstrate that the proposed approach is very effective, exhibiting a significantly better performance in selecting which subset of links to support compared to the existing convex approximation approaches.
%

\appendices

\section{Proof of Theorem \ref{thm-recover}}\label{app-recover}

To prove Theorem \ref{thm-recover}, we first introduce the following lemma.
\begin{yinli}[\cite{msp}]\label{lemma-admission}

   Suppose $\S$ is an admissible set\footnote{The subset $\S$ is called admissible if there exists a feasible $\bx$ such that $\left[\bA\bx-\bb\right]_{\S}\geq\mathbf{0}$.} of problem \eqref{sparse2} and ${\S}^c=\K\setminus\S$ is its complement.
                                                                Then the following statements hold.
  \begin{itemize}
\item [(1)] For any admissible set $\S,$ $\bA_{\S,\S}$ is invertible and $(\bA_{\S,\S})^{-1}\geq \bm{0}.$

\item [(2)] 
For any admissible set $\S$ and any feasible $\bx$ satisfying $\left[\bA\bx-\bb\right]_{\S}\geq\mathbf{0},$ 
\begin{equation}\label{loweradmission}\bx_{\S}\geq (\bA_{\S,\S})^{-1}\bb_{\S}>\bm{0}.\end{equation}
\item [(3)] Let $\bar\S$ be the optimal maximum admissible set of problem \eqref{sparse2}. Then $\bx^*$ with
\begin{equation}\label{xstar}\bx_{\bar\S}^*=(\bA_{\bar\S,\bar\S})^{-1}\bb_{\bar\S},~\bx_{{\bar\S}^c}^*=\mathbf{0}\end{equation} is the solution to problem \eqref{sparse2}.


  \end{itemize}
\end{yinli}



We are now ready to prove Theorem \ref{thm-recover}. We first {prove that} the theorem is true under the assumptions that the solution of problem \eqref{sparse2} is unique and the maximum admissible set of problem \eqref{sparse2} is also unique. Then, we remove these two assumptions and {prove that} the theorem remains true.

  Case I: Assume $\bx^*$ in \eqref{xstar} is the unique global minimizer of problem \eqref{sparse2} and $\bar\S$ is the corresponding maximum admissible set, which is also {unique}.  Next, we show $\bx^*$ is the unique global minimizer of problem \eqref{lp}. By Lemma \ref{balancing}, it is equivalent to show $\bx^*$ is the unique global minimizer of problem \eqref{qnorm-r}. We divide the proof into two parts.


  Part A: In this part, we show that when $q$ is sufficiently small, $\bx^*$ is the unique global minimizer of problem \begin{equation}\label{psub}
\begin{array}{cl}
\displaystyle \min_{\bx} & \displaystyle \|[\bb-\bA\bx]_{{\bar\S}^c}\|_q^q + \alpha\bar\bp^T\bx \\
[5pt] \mbox{s.t.} & [\bb-\bA\bx]_{\bar\S}=0,\\
& [\bb-\bA\bx]_{{\bar\S}^c}>0,\\
& \displaystyle \mathbf{0}\leq \bx\leq \be.
\end{array}
\end{equation}
  Consider the following problem
  \begin{equation}\label{lowerbound}
\begin{array}{ll}
\displaystyle \min_{\bx} & \displaystyle \min\left\{[\bb-\bA\bx]_{{\bar\S}^c}\right\}\\
[5pt] \mbox{s.t.} & [\bb-\bA\bx]_{{\bar\S}}=0,\\
& [\bb-\bA\bx]_{{\bar\S}^c}\geq 0,\\
& \displaystyle \mathbf{0}\leq \bx\leq \be.
\end{array}
\end{equation} We claim that the optimal value of problem \eqref{lowerbound} is greater than or equal to $\underline \delta:=\delta(\bA,\bb)>0.$ 
Otherwise, there must exist a feasible point $\bx$ of problem \eqref{lowerbound} such that
$$[\bb-\bA\bx]_{j}=0,~\forall~j\in{\bar\S}\cup k,$$ where $k\in\bar \S^c.$ This contradicts the fact that $\bar\S$ is the maximum admissible set. 
Suppose $\bg(\bx)$ is the gradient of the objective function in \eqref{psub}. Then 
  \begin{align*}
     ~\bg(\bx)=\alpha\bar\bp-q\left(\bA_{\bar\S^c,\bar\S},\bA_{\bar\S^c,\bar\S^c}\right)^T[\bb-\bA\bx]_{\bar\S^c}^{q-1}\geq \alpha\bar\bp-q\underline\delta^{q-1}\left(|\bA_{\bar\S^c,\bar\S}|,|\bA_{\bar\S^c,\bar\S^c}|\right)^T\be,
  \end{align*} where $|\bA|$ denotes the entry-wise absolute value of the matrix $\bA.$ If $q$ is sufficiently small (say, $q\leq \bar q_1(\bA,\bb,\alpha,\bar\bp),$ where $\bar q_1(\bA,\bb,\alpha,\bar\bp)$ is a positive number such that for any $q\in(0, \bar q_1(\bA,\bb,\alpha,\bar\bp)],$ we have $\alpha\bar\bp-q\underline\delta^{q-1}\left(|\bA_{\bar\S^c,\bar\S}|,|\bA_{\bar\S^c,\bar\S^c}|\right)^T\be>\mathbf{0}$), then the gradient $\bg(\bx)$ of the objective function in \eqref{psub} is component-wise positive at any feasible point. In addition, to guarantee $[\bb-\bA\bx]_{\bar\S}=\mathbf{0},$ we need $\bx_{\bar\S}\geq \bx_{\bar\S}^*=\bA_{\bar\S,\bar\S}^{-1}\bb_{\bar\S}$ (cf. \eqref{loweradmission} and \eqref{xstar}). Therefore, $\bx^*$ is the unique global minimizer of problem \eqref{psub}, since for any feasible $\bx,$ we have $\bx_{\bar\S}\geq \bx_{\bar\S}^*$ and $\bx_{\bar\S^c}\geq \mathbf{0}.$

  Part B: In this part, we show that when $q$ is sufficiently small, $\bx^*$ is the unique global minimizer of problem \eqref{qnorm-r}. To show this, it suffices to show that, for any given admissible set $\S\subset \K$ with $|{\S} |< |\bar\S|,$ the minimum value of problem \begin{equation}\label{psub2}
\begin{array}{cl}
\displaystyle \min_{\bx} & \displaystyle \|[\bb-\bA\bx]_{{\S}^c}\|_q^q + \alpha \bar\bp^T\bx \\
[5pt] \mbox{s.t.} & [\bb-\bA\bx]_{{\S}}=0,\\
& [\bb-\bA\bx]_{{\S}^c}>0,\\
& \displaystyle \mathbf{0}\leq \bx\leq \be,
\end{array}
\end{equation} is greater than the one of problem \eqref{psub}. Without loss of generality, suppose the solution $\bx^*({\S})$ of problem \eqref{psub2} is attainable. Otherwise, there must exist $k\in {\S}^c$ such that $[\bb-\bA\bx]_{k}\rightarrow0$ at the optimal point. In this case, we consider problem \eqref{psub2} with $\S$ replaced by ${\S}\bigcup k.$ According to the assumption that the maximum admissible set of problem \eqref{sparse2} is unique, we still have $|{\S}\bigcup k|<|\bar\S|$ unless ${\S}\bigcup k=\bar\S.$ 

Suppose $\bx^*({\S})$ is achievable, there must exist $\delta_{\S}>0$ such that
\begin{equation}\label{add}[\bb-\bA\bx^*({\S})]_j\geq \delta_{{\S}}>0,~\forall~j\in {\S}^c,\end{equation} where $\delta_{\S}$ depends on $\bA,\bb,\alpha,\bar\bp,\S.$
Since the number of admissible sets $\S,$ with which the solution $\bx^*(\S)$ of problem \eqref{psub2} is achievable, is finite, then
\begin{equation}\label{delta}\delta:=\min\left\{\min_{\S\subset\K} \left\{\delta_{\S}\right\},\min\left\{\bb/2\right\}\right\}>0.\end{equation} Here, $\delta$ only depends on $\bA,\bb,\alpha$ and $\bar\bp.$ Define
\begin{equation}\label{Delta}\Delta:=\max_{\mathbf{0}\leq \bx\leq \be,\,\bb-\bA\bx\geq \mathbf{0}}\|\bb-\bA\bx\|_{\infty}.\end{equation} 
Let ${\bar q}_2(\bA,\bb,\alpha,\bar\bp)$ be a positive number such that for any $q\in(0,{\bar q}_2(\bA,\bb,\alpha,\bar\bp)],$ we have
\begin{equation}\label{pequation}\left(K-|{\bar\S}|+1\right)\delta^q\geq\left(K-|{\bar\S}|\right)\Delta^q+\alpha\bar\bp^T\bx^*.\end{equation} 
Therefore, if \begin{equation}\label{barq}0< q\leq \bar q:=\min\left\{{\bar q}_1(\bA,\bb,\alpha,\bar\bp), {\bar q}_2(\bA,\bb,\alpha,\bar\bp)\right\},\end{equation} 
for any admissible $\S$ with $|{\S}|<|\bar\S|:$
\begin{itemize}
  \item if $\bx^*({\S})\neq \mathbf{0},$ there holds
  \begin{align}
  ~\|[\bb-\bA \bx^*({\S})]\|_q^q+\alpha\bar\bp^T\bx^*(\S)~>&~(K-|{\S}|)\delta^q\label{ineq1}\\
  \geq&~(K-|\bar\S|+1)\delta^q\label{ineq2}\\
  \geq&~\left(K-|{\bar\S}|\right)\Delta^q+\alpha\bar\bp^T\bx^*\label{ineq3}\\
  \geq&~\|\bb-\bA\bx^*\|_q^q+\alpha\bar\bp^T\bx^*,\label{ineq4}
\end{align}where \eqref{ineq1} is due to \eqref{add}, \eqref{delta}, and $\bx^*({\S})\neq 0,$ \eqref{ineq2} is due to the fact $|{\S}|<|\bar\S|$, \eqref{ineq3} is due to \eqref{pequation} and \eqref{barq}, and \eqref{ineq4} is by the definition of $\Delta$ (cf. \eqref{Delta}).
\item if $\bx^*({\S})= 0,$ then
$$\|\bb-\bA\bx^*({\S})\|_q^q+\alpha\bar\bp^T\bx^*({\S})=\|\bb\|^q> K\delta^q\geq\|\bb-\bA\bx^*\|_q^q+\alpha\bar\bp^T\bx^*,$$ where the first strict inequality is due to \eqref{delta}, and the second inequality can be obtained in a similar fashion as in the case of $\bx^*({\S})\neq \mathbf{0}.$
\end{itemize}


Case II: Consider the case when problem \eqref{sparse2} has multiple maximum admissible sets, but its solution $\bx^*$ remains unique. Then, for any feasible set $\S\neq \bar\S$ such that $|{\S}|=|\bar\S|,$ we have
$$\|\bA\bx^*({\S})-\bb\|_0+\alpha\bar\bp^T\bx^*({\S})>\|\bA\bx^*-\bb\|_0+\alpha\bar\bp^T\bx^*,$$ where $\bx^*(\S)$ is the solution to problem \eqref{psub2}. The above strict inequality is because $\bx^*$ is the unique solution to problem \eqref{sparse2}. Therefore, there exists $\bar q(\bA,\bb,\alpha,\bar\bp,{\S})>0$ such that for all $q\in(0, \bar q(\bA,\bb,\alpha,\bar\bp,{\S})],$ there holds
$$\|\bA\bx^*({\S})-\bb\|_q^q+\alpha\bar\bp^T\bx^*({\S})>\|\bA\bx^*-\bb\|_q^q+\alpha\bar\bp^T\bx^*.$$ Since problem \eqref{sparse2} has at most $\displaystyle {K \choose |\bar\S|}$ maximum admissible sets, we can take the minimum among $\left\{\bar q(\bA,\bb,\alpha,\bar\bp,{\S})\right\}$, and obtain a $\bar q(\bA,\bb,\alpha,\bar\bp)>0$ such that when $q\in(0,\bar q(\bA,\bb,\alpha,\bar\bp)],$ it has $$\|\bA\bx^*({\S})-\bb\|_q^q+\alpha\bar\bp^T\bx^*({\S})>\|\bA\bx^*-\bb\|_q^q+\alpha\bar\bp^T\bx^*,\forall~{\S}\neq\S^*,~|\S|=|\S^*|.$$ This, together with Case I, implies that when $q$ is sufficiently small, $\bx^*$ is the unique global minimizer of problem \eqref{qnorm-r} under the assumption that the solution $\bx^*$ of problem \eqref{sparse2} is unique.

Case III: The remaining case is that when problem \eqref{sparse2} has multiple solutions. Without loss of generality, we assume that there are two different solutions $\bx_1^*$ and $\bx_2^*.$
Then, the choice of $\alpha$ (cf. \eqref{alpha1}) immediately implies $$\|\bA\bx_1^*-\bb\|_0=\|\bA\bx_2^*-\bb\|_0,~\bar\bp^T\bx_1^*=\bar\bp^T\bx_2^*.$$
If 
there exists an bijective mapping $\pi$ from $\left\{1,2,\ldots,K\right\}$ to $\left\{1,2,\ldots,K\right\}$ such that $[\bA\bx_1^*-\bb]_j=[\bA\bx_2^*-\bb]_{\pi(j)}$ for all $j\in\K,$ then both $\bx_1^*$ and $\bx_2^*$ are global minimizers of problem \eqref{qnorm-r}. Otherwise, we can find $\bar q(\bA,\bb,\alpha,\bar\bp)>0$ such that when $q\in(0,\bar q(\bA,\bb,\alpha,\bar\bp)],$ we have either
$$\|\bA\bx_1^*-\bb\|_q^q+\alpha\bar\bp^T\bx_1^*<\|\bA\bx_2^*-\bb\|_q^q+\alpha\bar\bp^T\bx_2^*$$ or
 $$\|\bA\bx_1^*-\bb\|_q^q+\alpha\bar\bp^T\bx_1^*>\|\bA\bx_2^*-\bb\|_q^q+\alpha\bar\bp^T\bx_2^*.$$
Combining the above with Cases I and II, we know that $\bx_1^*$ (or $\bx_2^*$) is the global minimizer of problem \eqref{qnorm-r}. This completes the proof of Theorem \ref{thm-recover}.
%
%
%

\section{Proof of Theorem \ref{hardness}}\label{app-hardness}
Given an instance of the partition problem with $s_1,s_2,\ldots,s_N,$ define $S=\sum_{n=1}^Ns_n.$ Next, we construct an instance of problem \eqref{lp}, where 
\begin{itemize}
\item $K=2N+2;$
    \item all entries of $\bar\bp\in\mathbb{R}^{2N+2}$ are set to be $1;$
  \item the first $2N$ entries of $\bb\in\mathbb{R}^{2N+2}$ are set to be $1,$ and the last two entries of $\bb$ are set to be $0.5;$ 
  \item all diagonal entries of $\bA\in \mathbb{R}^{\left(2N+2\right)\times\left(2N+2\right)}$ are one, and non-diagonal entries of $\bA$ are 
  \begin{itemize}
    \item [-] for $k=2n-1,~n=1,2,\ldots,N,$ set $a_{k,j}=0$ except $a_{k,k+1}=-1;$
    \item [-] for $k=2n,~n=1,2,\ldots,N,$ set $a_{k,j}=0$ except $a_{k,k-1}=-1;$
    \item [-] for $k=2N+1,$ set $a_{k,j}=0$ except $a_{k,2n-1}=-\frac{s_n}{S}$ for $n=1,2,\ldots,N;$
    \item [-] for $k=2N+2,$ set $a_{k,j}=0$ except $a_{k,2n}=-\frac{s_n}{S}$ for $n=1,2,\ldots,N;$ and

  \end{itemize}
  \item the parameter $\alpha$ satisfies\begin{equation}\label{aq}0<\alpha\leq\frac{2-2^q}{2}.
  \end{equation}
\end{itemize}
%
Then the constructed instance of problem \eqref{lp} is
\begin{equation}\label{instance}
\begin{array}{cl}
\displaystyle \min_{x_1,...,x_{2N+2}} & \displaystyle \sum_{n=1}^NF(x_{2n-1}, x_{2n})+\sum_{i=0}^1H(x_{1+i},x_{3+i},\ldots,x_{2N+1+i})\\
\mbox{s.t.} & \displaystyle {0}\leq x_n \leq 1,~n=1,2,\ldots,2N+2,
\end{array}
\end{equation}where
\begin{align*}
&F(x_{2n-1},x_{2n})=~\left|1+x_{2n}-x_{2n-1}\right|^q+\left|1+x_{2n-1}- x_{2n}\right|^q+\alpha(x_{2n-1}+x_{2n}),~n=1,2,\ldots,N,\\[3pt]
&H(x_{1+i},x_{3+i},\ldots,x_{2N+1+i})=~\left|0.5+\frac{1}{S}\sum_{n=1}^Ns_nx_{2n-1+i}-x_{2N+1+i}\right|^q+\alpha x_{2N+1+i},~i=0,1.
\end{align*} Notice that $0\leq x_{2n-1}, x_{2n}\leq 1$ for all $n=1,2,\ldots,N,$ it follows that
\begin{equation}\label{Fx}F(x_{2n-1}, x_{2n})=\left(1+x_{2n}-x_{2n-1}\right)^q+\left(1+x_{2n-1}-x_{2n}\right)^q+\alpha(x_{2n-1}+x_{2n}),~n=1,2,\ldots,N.\end{equation}

Next, we claim that the partition problem has a ``yes'' answer if and only if the optimal value of problem \eqref{instance} is less than or equal to { $2^qN+\alpha (N+2).$}  We prove the ``if'' and ``only if'' directions separately.

Let us first prove the ``only if'' direction. Suppose the partition problem has a ``yes'' answer and let $\S$ be the subset of $\left\{1,2,\ldots,N\right\}$ such that \begin{equation}\label{partition}\sum_{n\in\S}s_n={S}/{2}.\end{equation} We show that {there exists} a feasible power allocation vector $\left\{x_n\right\}_{n=1}^{2N+2}$ such that the optimal value of problem \eqref{instance} is less than or equal to $2^q N+\alpha (N+2).$ In particular, let
\begin{equation}\label{sol}\hat x_{2n-1}=\left\{\begin{array}{ll}
       1, & \text{if}~n\in\S;\\
       0, & \text{if}~n\notin\S,
       \end{array}\right.\quad ~\hat x_{2n}=1-\hat x_{2n-1},~n=1,2,\ldots,N,~\end{equation} and $$\hat x_{2N+1}=\hat x_{2N+2}=1.$$ It is simple to check 
       \begin{align}
       F(\hat x_{2n-1},\hat x_{2n})&=~2^q+\alpha,~n=1,2,\ldots,N,\nonumber\\
       H(\hat x_1,\hat x_3,\ldots,\hat x_{2N+1})& =~\left|0.5+\frac{1}{S}\sum_{n=1}^Ns_n\hat x_{2n-1}-\hat x_{2N+1}\right|^q+\alpha \hat x_{2N+1}\nonumber\\
               &=~\left|0.5+\frac{1}{S}\sum_{n\in\S}s_n-1\right|^q+\alpha\quad
        (\text{from}~\eqref{sol})\nonumber\\
               &=~\alpha,\quad
        (\text{from}~\eqref{partition})\nonumber\\
        H(\hat x_2,\hat x_4,\ldots,\hat x_{2N+2})& =~\alpha.\nonumber
       \end{align}
     Thus, we have
$$\sum_{n=1}^NF(\hat x_{2n-1},\hat x_{2n})+\sum_{i=0}^1H(\hat x_{1+i},\hat x_{3+i},\ldots,\hat x_{2N+1+i})=N2^q+(N+2)\alpha,$$ which implies that the optimal value of problem \eqref{instance} is less than or equal to $N 2^q+(N+2)\alpha.$
 %

To show the ``if'' direction, suppose that the optimal solution of problem \eqref{instance} is less than or equal to $N 2^q+(N+2)\alpha.$ 
%
%
%
Consider a relaxation of problem \eqref{instance} by dropping the constraints $x_{2N+1}\leq 1$ and $x_{2N+2}\leq 1:$
\begin{equation}\label{relaxation}
\begin{array}{cl}
\displaystyle \min_{{x_1,...,x_{2N+2}}} & \displaystyle \sum_{n=1}^NF(x_{2n-1}, x_{2n})+\sum_{i=0}^1H(x_{1+i},x_{3+i},\ldots,x_{2N+1+i})\\
\mbox{s.t.} & \displaystyle {0}\leq x_n\leq 1,~n=1,2,\ldots,2N,\\
            & x_{2N+1}\geq 0,~x_{2N+2}\geq 0.
\end{array}
\end{equation}Clearly, the optimal value of problem \eqref{relaxation} is less than or equal to the optimal value of problem \eqref{instance}.
The relaxed problem \eqref{relaxation} can be equivalently rewritten as
\begin{equation*}
\begin{array}{cl}
\displaystyle \min_{x_1,...,x_{2N+2}} & \displaystyle \sum_{n=1}^NF(x_{2n-1}, x_{2n})+\sum_{i=0}^1\hat H(x_{1+i},x_{3+i},\ldots,x_{2N-1+i})\\
\mbox{s.t.} & \displaystyle {0}\leq x_n\leq 1,~n=1,2,\ldots,2N,
\end{array}
\end{equation*}
where {for $i=0, 1,$} \begin{equation}\label{sub1}
\begin{array}{rcl}
\hat H(x_{1+i},x_{3+i},\ldots,x_{2N-1+i}):=\!\!\!&\displaystyle \min_{x_{2N+1+i}} & H(x_{1+i},x_{3+i},\ldots,x_{2N-1+i},x_{2N+1+i}) \\
&\mbox{s.t.} & \displaystyle x_{2N+1+i}\geq 0.
\end{array}
\end{equation}
Since problem \eqref{sub1} is an univariate optimization problem and {$\alpha$ satisfies \eqref{aq}}, we can verify that, for any $x_{1+i},x_{3+i},\ldots,x_{2N+1+i}\geq 0,$ there holds
\begin{equation}\label{minimum}H(x_{1+i},x_{3+i},\ldots,x_{2N+1+i})\geq H(x_{1+i},x_{3+i},\ldots,\hat x_{2N+1+i})=\frac{\alpha}{2}+\frac{\alpha}{S}\sum_{n=1}^{N}s_nx_{2n-1+i},\end{equation} where 
\begin{equation}\label{hatx}\hat x_{2N+1+i}=0.5+\frac{1}{S}\sum_{n=1}^Ns_nx_{2n-1+i},~i=0,1.\end{equation}
By definition \eqref{sub1}, we have
$$\hat H(x_{1+i},x_{3+i},\ldots,x_{2N-1+i})=\frac{\alpha}{2}+\frac{\alpha}{S}\sum_{n=1}^{N}s_nx_{2n-1+i},~i=0,1.$$
As a result, problem \eqref{relaxation} {can be decomposed} into $N$ subproblems
\begin{equation}\label{np-sub2}
\begin{array}{cl}
\displaystyle \min_{x_{2n-1},x_{2n}} & \hat F(x_{2n-1}, x_{2n}):=F(x_{2n-1}, x_{2n})+\frac{\alpha s_n}{S}\left(x_{2n-1}+x_{2n}\right) \\
\mbox{s.t.} & \displaystyle {0}\leq x_{2n-1},\, x_{2n}\leq 1.
\end{array}
\end{equation}
  We know from \eqref{Fx} that $\hat F(x_{2n-1},x_{2n})$ in \eqref{np-sub2} is strictly concave with respect to $x_{2n-1}$ and $x_{2n}$ in $[0,1]\times[0,1].$ Since the minimum of a strictly concave function is always attained at a vertex\cite{concave}, we immediately obtain that the optimal solution of \eqref{np-sub2} must be $(0,0),$ $(0,1),$ $(1,0),$ or $(1,1).$ It is easy to see that $$\hat F(0,0)=2,~\hat F(0,1)=\hat F(1,0)=2^q+\alpha\left(1+\frac{s_n}{S}\right),~\hat F(1,1)=2+2\alpha\left(1+\frac{s_n}{S}\right).$$
This, together with the facts $0<s_n/S<1$ and $\alpha\leq\frac{2-2^q}{2}$ (cf. \eqref{aq}), shows the optimal solution of \eqref{np-sub2} is \begin{equation}\label{sol2}(\hat x_{2n-1},\hat x_{2n})=(0,1)~\text{or}~(1,0).\end{equation} Now, we can use \eqref{hatx} and \eqref{sol2} to conclude that the optimal value of problem \eqref{relaxation} is $$\sum_{n=1}^N \left(2^q+\alpha+\frac{{\alpha} s_n}{S}\right)+\alpha=N2^q+(N+2)\alpha.$$ Since the optimal value of problem \eqref{instance} is less than or equal to $N2^q+(N+2)\alpha$ (the assumption of the ``if'' direction), it follows from \eqref{hatx} that 
$$\hat x_{2N+1}=0.5+\frac{1}{S}\sum_{n=1}^Ns_n\hat x_{2n-1}\leq 1,~\hat x_{2N+2}=0.5+\frac{1}{S}\sum_{n=1}^Ns_n\hat x_{2n}\leq 1.$$
Combinging this with \eqref{sol2} yields
$$\sum_{n=1}^Ns_n\hat x_{2n-1}=\sum_{n\in\S}s_n=\frac{S}{2},~\sum_{n=1}^Ns_n\hat x_{2n}=\sum_{n\notin\S}s_n= \frac{S}{2},$$ where ${\S}=\left\{n\,|\,\hat x_{2n-1}=1\right\}.$ Therefore, there exists a subset $\S$ such that \eqref{partition} holds true, which shows that the partition problem has a ``yes'' answer.

Finally, this transformation can be finished in polynomial time. Since the partition problem is NP-complete, we conclude that problem \eqref{lp} is NP-hard.

\section{Proof of Lemma \ref{lemma-lowerbound}}\label{app-potential}
Since $\bw$ is feasible,
  ~it follows that
\begin{equation}\label{in1}\sum_{k=1}^K\log([\bw_1]_k[\bw_3]_k)=\sum_{k=1}^K\log([\bw_1]_k(1-[\bw_1]_k))\leq -K\log(4),\end{equation} and
\begin{equation}\label{in2}
  \frac{K}{q}\log(\|\bw_2\|_q^q)-\sum_{k=1}^K\log([\bw_2]_k)\geq \frac{K}{q}\log K,
\end{equation}where \eqref{in2} comes from
$$\displaystyle \frac{\|\bw_2\|_q^q}{K}=\frac{\sum_{k=1}^K [\bw_2]_k^q}{K}\geq \left(\displaystyle\Pi_{k=1}^K [\bw_2]_k^q\right)^{1/K}.$$
By the definition of $\phi(\bw)$ (cf. \eqref{potential}), we obtain, for any strictly feasible $\bw,$
\begin{equation}\label{bigeq}\begin{array}{rl}
  \phi(\bw)=&\!\!\!\rho\log (f(\bw))-\sum_{k=1}^K\log([\bw_1]_k)-\sum_{k=1}^K\log([\bw_2]_k)-\sum_{k=1}^K\log([\bw_3]_k)\\[5pt]
             \geq&\!\!\!\left(\rho-\frac{K}{q}\right)\log(f(\bw))+\left(\frac{K}{q}\log(\|\bw_2\|_q^q)-\sum_{k=1}^K\log([\bw_2]_k)\right)-\sum_{k=1}^K\log([\bw_1]_k[\bw_3]_k) \\[10pt]
             \geq&\!\!\!\left(\rho-\frac{K}{q}\right)\log(f(\bw))+\frac{K}{q}\log \left(K\right)+K\log(4), 
\end{array}\end{equation}where the first inequality is because $\tilde\bc^T\bw_1\geq0$ for any feasible $\bw,$ and the second is by \eqref{in1} and \eqref{in2}.
Therefore, if \eqref{pupper} holds,~
we must have $f(\bw)\leq \epsilon,$ which shows that $\bw$ is an $\epsilon$-optimal solution.

\section{Proof of Theorem \ref{theoremK}}\label{app-polynomialtimecomplexity}
To show the polynomial time complexity of the potential reduction algorithm, we consider the following two cases.
 \begin{itemize}
   \item If $\|g(\bw)\|>1,$ then we know $$\phi(\bw^+)-\phi(\bw)<-\beta +\frac{\beta^2}{2(1-\beta)}.$$ The potential function value is reduced by a constant $2-\sqrt{3}$ if we set $\beta=1-\frac{\sqrt{3}}{3}<1.$ If this case would hold for $O\left(\left(\rho-\frac{K}{q}\right)\log\left(\frac{1}{\epsilon}\right)\right)$ iterations (cf. \eqref{pupper} and \eqref{upperbound-w}), we would obtain an $\epsilon$-optimal solution of \eqref{pnorm1}.
   \item If $\|g(\bw)\|\leq 1,$ then, from the definition of $g(\bw),$ we must have
$$0\leq \frac{\rho}{f(\bw)}\bW(\nabla f(\bw)-\tilde \bA^T\bm{\lambda})\leq 2\be.$$ In other words,

$$\left[\nabla f(\bw)-\tilde \bA^T\bm{\lambda}\right]_k\geq \bm{0},~\frac{\bw_k\left[\nabla f(\bw)-\tilde \bA^T\bm{\lambda}\right]_k}{f(\bw)}\leq \frac{2}{\rho},~\forall~k.$$ By choosing $\rho\geq \frac{6K}{\epsilon},$ we have $$\frac{\bw^T(\nabla f(\bw)-\tilde \bA^T\bm{\lambda})}{f(\bw)}\leq\epsilon.$$ Therefore,
$$\displaystyle\frac{\sum_{n=1}^{3K}\left(q\left[\bw\right]_n^{q} -\left[\tilde\bA^T\bm{\lambda}\right]_n\left[\bw\right]_n\right)}{\bar f-\underline f}=\frac{\bw^T(\nabla f(\bw)-\tilde \bA^T\bm{\lambda})}{\bar f-\underline f}\leq \frac{\bw^T(\nabla f(\bw)-\tilde \bA^T\bm{\lambda})}{f(\bw)}\leq\epsilon.$$
Recalling Definition \ref{eKKT}, we know that $\bw$ is an $\epsilon$-KKT point of problem \eqref{pnorm1}. 
 \end{itemize}

\section*{Acknowledgment}
The authors would like to thank Professor Zhi-Quan (Tom) Luo of University of Minnesota, Professor Shuzhong Zhang of University of Minnesota, and Dr. Qingna Li of Beijing Institute of Technology for their useful discussions. 

 \end{document}